\documentclass[11pt,a4paper]{article}
\usepackage{jheppub}
\usepackage[utf8]{inputenc}
\usepackage{graphicx,fancybox,float,comment}
\usepackage{units}
\usepackage{multirow,array,arydshln}
\usepackage[dvipsnames]{xcolor}
\usepackage{braket,tensor}
\usepackage{amsmath,amssymb,amsfonts}
\usepackage{tikz-cd}

\usepackage{ulem}

\global\interfootnotelinepenalty=1000

\title{
Chaos and pole-skipping in a simply spinning plasma }

\author[1,3]{Markus A.G. Amano,} 
\emailAdd{markus@henu.edu.cn}
\author[2]{Mike Blake,}
\emailAdd{mike.blake@bristol.ac.uk}
\author[3,4]{Casey Cartwright,} 
\emailAdd{c.c.cartwright@uu.nl}
\author[3]{Matthias Kaminski}
\emailAdd{mski@ua.edu}
\author[2,5]{and Anthony P. Thompson}
\emailAdd{at16718@bristol.ac.uk}
\affiliation[1]{Institute of Contemporary Mathematics, School of Mathematics and Statistics, Henan University, Kaifeng, Henan 475004, P. R. China}
\affiliation[2]{School of Mathematics, Fry Building, Woodland Road, Bristol BS8 1UG, U.K}
\affiliation[3]{Department of Physics and Astronomy, University of Alabama,\\ Tuscaloosa, AL 35487, USA}
\affiliation[4]{Institute for Theoretical Physics, Utrecht University, Princetonplein 5, 3584 CC Utrecht, The Netherlands}
\affiliation[5]{Quantum Engineering Center for Doctoral Training, University of Bristol, UK}

\abstract{We study the relationship between many-body quantum chaos and energy dynamics in holographic quantum field theory states dual to the simply-spinning Myers-Perry-AdS$_5$ black hole. The enhanced symmetry of such black holes allows us to provide a thorough examination of the phenomenon of pole-skipping, that is significantly simpler than  a previous analysis of quantum field theory states dual to the Kerr-AdS$_4$ solution. In particular we give a general proof of pole-skipping in the retarded energy density Green's function of the dual quantum field theory whenever the spatial profile of energy fluctuations satisfies the shockwave equation governing the form of the OTOC. Furthermore, in the large black hole limit we are able to obtain a simple analytic expression for the OTOC for operator configurations on Hopf circles, and demonstrate that the associated Lyapunov exponent and butterfly velocity are robustly related to the locations of a family of pole-skipping points in the energy response. Finally, we note that in contrast to previous studies, our results are valid for any value of rotation and we are able to numerically demonstrate that the dispersion relations of sound modes in the energy response explicitly pass through our pole-skipping locations.}

\begin{document}

\maketitle

\section{Introduction}
\paragraph{} Holographic quantum field theories provide an important source of tools and insight into the physics of strongly coupled systems. One such important insight has been the phenomenon of pole-skipping which identifies special points in complex momentum space $(\omega, k)$ where the retarded Green's function becomes ill-defined \cite{2018G, 2018B, 2018, 2020, Grozdanov_2019}. Remarkably, it has been quite generally established that in maximally chaotic systems there exists a family of pole-skipping points in the energy density response  $(\omega^*, k^*)$ whose locations are robustly related to the Lyapunov exponent and butterfly velocity that govern the form of out-of-time-ordered correlation functions (OTOCs). This provides a remarkable connection between energy dynamics and quantum many-body chaos in such systems, and supports a proposal of \cite{2018B, Blake:2021wqj} for a hydrodynamic origin of chaos in such systems. 

\paragraph{} Since the discovery of the pole-skipping phenomenon, there has been much work studying it in static planar black holes \cite{2018G, 2018B, 2018, 2020, Grozdanov_2019, Grozdanov2_2019, Natsuume_2020, Wang:2022mcq}. More recent work has studied the connection between chaos and energy-dynamics in rotating black holes. The simplest case is that of the rotating BTZ black hole, which is dual to a $1+1$ dimensional conformal field theory with chemical potential for rotation, which was studied in detail in ref. \cite{Liu_2020} (see also \cite{2020Mezei, 2019Jahnke} for the derivation  of the form of the OTOC).  Chaos and pole-skipping in the field theory dual to 3+1 dimensional  rotating Kerr-AdS black hole was recently studied in \cite{blake2021chaos}. While the phenomenon of pole-skipping was still found to hold in this example, the analysis is extremely technical and sophisticated. As a result of this, there are some limitations to the analysis carried out in ref. \cite{blake2021chaos}. Firstly, it was only possible to study the form of the OTOC in the slowly rotating limit (limit of small chemical potential). Secondly, though an analytical argument was given that showed the existence of a quasinormal mode at the pole-skipping frequencies, this was not verified numerically. Thirdly, since the linearized Einstein's equations are not separable in the Kerr-AdS geometry it is only possible to verify the pole-skipping phenomenon using the Teukolsky formalism. This makes it valuable to provide an example of pole-skipping in the field theory dual to a higher dimensional rotating black hole which is more tractable to study than the Kerr-AdS geometry, which is the aim of this paper.\\

A spinning object in $4+1$D has two orthogonal planes of rotation, so is characterised by two angular momenta instead of one. The Myers-Perry-AdS geometry generalizes the Kerr metric to higher dimensions and provides the geometry for a spinning black hole in $4+1$D \cite{myers_perry_1986}. The isometry group of the Myers-Perry-AdS geometry is $\mathbb{R}\times U(1)^2$, with a symmetry for time translation invariance and two copies of $U(1)$ for each plane of rotation. However, in the case where the two angular momenta are set equal to each other the geometry has an enhanced symmetry group $\mathbb{R}\times SU(2)\times U(1)$ and the metric becomes co-homogeneity-1, so that it depends non-trivially only on the radial parameter. This simplified geometry presents a natural higher dimensional geometry to study the phenomenon of pole-skipping.\\

In Section~\ref{MP} we will review the details of the geometry of the Myers-Perry-AdS black hole with equal angular momenta. In Section~\ref{OTOC} we study the OTOC using the eikonal phase approximation, and derive the equation governing the angular profile of a gravitational shockwave\footnote{The effects of axisymmetrical gravitational shockwaves on mutual information in the Myers-Perry-AdS$_5$ black hole were recently studied in \cite{Malvimat:2022fhd}.}  at the horizon which is associated with a high energy scattering process \cite{shenker2015stringy}. We refer to the equation governing the angular profile as the shockwave equation, and we find that it can be solved analytically using Wigner D-functions. The most tractable regime is the large black hole limit, where the OTOC on Hopf circles of the $3$-sphere takes a simple form. A key part of the results of this paper is that we are able to obtain an analytic solution to the shockwave equation that determines the form of the OTOC in large black holes away from the slowly rotating limit.\\

Following this we study the energy-density response and identify the pole-skipping points. In Section~\ref{NH} we study ingoing metric perturbation near the horizon, and demonstrate the energy response of the dual boundary theory exhibits pole-skipping at $\omega = i 2 \pi T$\footnote{This is the frequency in co-rotating angular coordinates.} when the angular profile of perturbations satisfies the sourceless shock-wave equation discussed in Section~\ref{OTOC}. In particular, for metric perturbations of frequency $\omega = i2\pi T$ the $vv$-component of the linearized Einstein equations becomes identical to the (sourceless) shockwave equation, ensuring the existence of an extra ingoing mode for perturbations whose angular profile satisfies this equation. Due to the enhanced symmetry of the Myers-Perry-AdS black hole with equal angular momenta the Einstein equations separate into a coupled system of ODE's. This ensures that the existence of an additional solution to the linearized Einstein equations near the horizon will guarantee the existence of an additional ingoing mode everywhere in the bulk, which is the gravitational origin of pole-skipping as first shown in~\cite{2018B}.

\paragraph{} Again the most tractable limit corresponds to the large black hole limit. In this limit we demonstrate that for perturbations whose angular momentum is aligned with the rotation of the black hole the complex frequencies and wave-vectors associated to the pole-skipping points precisely match those extracted from the OTOC for operator configurations on Hopf circles, in an analogous manner to static black holes. In Section~\ref{Num}, we perform a more detailed analysis of these pole-skipping points. In particular we demonstrate that they are equivalent to those of a boosted black brane, and numerically confirm using two distinct methods that the dispersion relations of sound modes in the energy response of the dual theory pass through these pole-skipping points. Finally, we emphasise that in contrast to previous studies of higher dimensional rotating black holes in \cite{blake2021chaos}, all of our results are valid for any value of the rotation parameter $a/L$. This allows us to identify several new physical aspects of chaos and pole-skipping in rotating systems. In particular, these include the observation that in the rest frame of the boundary theory one of the pole-skipping points associated to chaos crosses into the lower half of the complex frequency plane when the velocity $a/L$ associated to the rotation of a large black hole exceeds the conformal value of the butterfly velocity.

\section{Myers-Perry-AdS$_{5}$ black hole}\label{MP}
\paragraph{} We begin by reviewing the simply spinning Myers-Perry-AdS$_{5}$ {\cite{Chen:2006xh, Gibbons:2004js, Murata:2008xr, Hawking:1998kw}} black hole, and introducing the coordinate systems that will be necessary to study chaos and pole-skipping in the dual quantum field theory. In particular, the Myers-Perry-AdS$_{5}$ black hole for equal choices of angular momenta can be described by the metric
\begin{equation}
    \label{metric}
    ds^2 = -\left(1 + \frac{r^2}{L^2}\right)dt^2 + \frac{dr^2}{G(r)} + \frac{r^2}{4}\left( \sigma_1^2 + \sigma_2^2 + \sigma_3^2\right) + \frac{2\mu}{r^2}\left(dt + \frac{a}{2}\sigma_3\right)^2, 
    \end{equation} 
where 
\begin{equation} 
G(r) = 1 +\frac{r^2}{L^2} - \frac{2\mu(1-a^2/L^2)}{r^2} + \frac{2\mu a^2}{r^4}.  \nonumber 
\end{equation}
The above metric has two horizons at the zeros of $G(r)$. The mass constant $\mu$ is related to the AdS-radius, $L$, the rotation parameter, $a$, and the location of the outer horizon, $r_0$, by
\begin{equation}
 \mu = \frac{r_0^4(L^2 + r_0^2)}{2L^2r_0^2 - 2a^2(L^2 + r_0^2)}. 
\end{equation}
The one-forms that appear in the metric \eqref{metric} are given by:
\begin{align*}
    \sigma_1 &= \mathrm{sin}(\theta) \mathrm{cos}(\psi) d\phi - \mathrm{sin}(\psi)d\theta, \\
    \sigma_2 &= \mathrm{cos}(\psi)d\theta + \mathrm{sin}(\theta)\mathrm{sin}(\psi)d\phi,  \\
    \sigma_3 &= d\psi + \mathrm{cos}(\theta)d\phi,
\end{align*}
where the angular coordinates $(\psi, \phi, \theta)$ parameterise a $3-$sphere in terms of the Hopf fibration, with $(\phi, \theta)$ being coordinates on the base $S^2$ and $\psi$ the coordinate on the fiber. Their ranges are given by $0 \leq \theta < \pi$, $0\leq \phi < 2\pi$ and $0\leq \psi < 4\pi$ \cite{Murata:2008xr}. The variable $a$ parameterises the simultaneous rotation of the black hole in two perpendicular planes of rotation\footnote{The angular coordinates $(\psi, \phi, \theta)$ are related to the embedding of the 3-sphere in $\mathbb{R}^{4}$ by $x_1 = \mathrm{cos}((\psi - \phi)/2) \mathrm{sin}(\theta/2)$, $ x_2 = \mathrm{sin}((\psi - \phi)/2) \mathrm{sin}(\theta/2)$, $x_3 = \mathrm{sin}((\psi + \phi)/2) \mathrm{cos}(\theta/2)$,  $x_4 = \mathrm{cos}((\psi + \phi)/2)\mathrm{cos}(\theta/2)$.} in $\mathbb{R}^4$. In the above coordinates, this can be understood as a rotation in the direction of the $\psi$ coordinate. The metric \eqref{metric} solves Einstein's equations with negative cosmological constant $\Lambda = -6/L^2$. 

\paragraph{} It will prove convenient to introduce some compact notation for various combinations of the constant parameters $r_0, a, L$ that will appear throughout this paper. We therefore define constant parameters $\Sigma, \Delta$ and $K$ by 
\begin{equation}
\Sigma = L^2 + r_0^2,  \hspace{1cm} \Delta = L^2r_0^2(L^2 + 2r_0^2), \hspace{1cm} K = \sqrt{L^2r_0^2 - a^2(L^2 + r_0^2)}. \nonumber
\end{equation}
In terms of these parameters the temperature of the dual black hole is then given by {\cite{Murata:2008xr, Gibbons:2004ai}}
\begin{equation}
\label{temperature}
2\pi T =  \frac{L^2r_0^2(L^2 +2r_0^2) - 2a^2(L^2 + r_0^2)^2}{L^3r_0^2\sqrt{L^2r_0^2 - a^2(L^2 + r_0^2)}} = \frac{\Delta -2 a^2 \Sigma^2}{L^3 r_0^2 K},
\end{equation}
and the angular velocity\footnote{An extremal solution occurs when $a/L=\pm\frac{\sqrt{L^2 r_0^2+2 r_0^4}}{\sqrt{2} \sqrt{L^4+2 L^2 r_0^2+r_0^4}} $ and consequently the temperature vanishes. In the large black hole limit this expression reduces to $a/L=\pm 1$. } of the outer-horizon is given by:
\begin{equation}
\label{omega}
\Omega = -2a\left(\frac{1}{L^2} + \frac{1}{r_0^2}\right). 
\end{equation}
The dual quantum field theory corresponds to a conformal field theory on $S^3 \times R^{t}$, at finite temperature \eqref{temperature} and with equal chemical potentials for rotation in two orthogonal planes. 

\paragraph{}The near horizon structure of the geometry is not clear from the form of the metric \eqref{metric}. In order to elucidate it, and introduce the analogue of Eddington-Finkelstein and Kruskal coordinates, we first introduce the co-rotating angular coordinate $\tilde{\psi}$ and the associated one form $\tilde{\sigma}_3$ by
\begin{equation}
    \tilde{\psi} = \psi - \Omega t, \hspace{2cm}  \tilde{\sigma}_3 = d\tilde{\psi} + \mathrm{cos}(\theta)d\phi,
    \label{co-rot}
\end{equation}
with $\Omega$ defined in \eqref{omega}. Then in co-rotating coordinates $(t, r, \tilde{\psi}, \phi, \theta)$ the $(t,r)$ components of the metric \eqref{metric} take the familiar form 
\begin{equation}
ds_{(t,r)}^2 = - F(r) dt^2 + \frac{dr^2}{G(r)},
\label{metricrt}
\end{equation}
where $F(r)$ is given by 
\begin{equation}
F(r) = 1 + \frac{r^2}{L^2}  - \frac{r_0^2}{r^2}\bigg(1 + \frac{r_0^2}{L^2}\bigg) - \frac{a^2 \Sigma^2 (r^4 -r_0^4)}{L^4 r^2 r_0^4},
\end{equation}
which vanishes at $r=r_0$ as expected in co-rotating coordinates\footnote{Note the temperature~\eqref{temperature} can be extracted as usual from~\eqref{metricrt} as $4 \pi T = H'(r_0)$ with $H(r) = \sqrt{F(r) G(r)}$.}. 

\paragraph{} We can now introduce analogues of ingoing ($v$) and outgoing ($u$) Eddington-Finkelstein coordinates by:
\begin{align}
    \label{ingoing}
    dv = dt + \frac{dr}{\sqrt{F(r)G(r)}}, \\
    du = dt - \frac{dr}{\sqrt{F(r)G(r)}}. \nonumber
\end{align}
Finally, we can proceed as usual to define Kruskal-like coordinates, $U$ and $V$,which will be necessary for studying the OTOC.  Defining 
\begin{equation}
\label{kruskdef}
    U = -e^{-\alpha u}, \hspace{2cm} V = e^{\alpha v}, 
\end{equation}
where $\alpha = 2 \pi T$, we find the metric in Kruskal coordinates is given by 
\begin{equation}
    ds^2  = A(UV)  dUdV + B(U V)(UdV - VdU)\tilde{\sigma}_3 + \frac{r^2}{4}(\sigma_1^2 + \sigma_2^2 +\tilde{\sigma}_3^2) + \frac{\mu a^2}{2 r^2}\tilde{\sigma}_3^2
    \label{Kruskal}
\end{equation}
where the function $r = r(UV)$ is defined implicitly through \eqref{kruskdef} and 
\begin{equation}
A(UV) =  \frac{F(r)}{\alpha^2 UV}, \;\;\;\;\;\;\;\;  B(U V) = \frac{a}{2 \alpha U V} \frac{\Sigma(r_0^4 - r^4)}{L^2 r^2 r_0^2}. 
\end{equation}
The metric \eqref{Kruskal} is manifestly smooth at both the $U=0$ and $V=0$ horizons at which $r(0)=r_0$.

\section{The OTOC for Myers-Perry-AdS$_5$ black holes}
\label{OTOC}

\paragraph{} We now wish to examine the computation of out-of-time ordered correlators (OTOCs) in the field theory dual to the geometry described by (\ref{metric}). As is now well known, for any holographic theory dual to classical gravity this can be computed by adapting the eikonal approximation to gravitational scattering presented in \cite{shenker2015stringy}. In particular, we will derive the shock-wave equation governing that angular profile of gravitational shockwaves that govern the form of the OTOC. We will see that resulting shockwave equation is significantly simpler that the one previously obtained for the Kerr-AdS black hole in \cite{blake2021chaos}. In particular it is possible to obtain an analytic solution for the shockwave profile as an expansion in Wigner D-functions, which provide analogues of spherical harmonics for the three-sphere. Furthermore, the computation of the OTOC significantly simplifies in the large black hole limit, which allows us to obtain a simple closed form expression for the OTOC for operators lying on a Hopf circle for any value of the rotation strength. This is in contrast to the case of the Kerr-AdS black studied in \cite{blake2021chaos}, for which exact results for the OTOC were only presented in the equatorial plane and in the slowly rotating limit of large black holes. 

\paragraph{} In particular, for concreteness we consider the OTOC 
\begin{equation}
    \langle \hat{W}(t_2, \Psi_2, \Phi_2, \Theta_2) \hat{V}(t_1, \Psi_1, \Phi_1, \Theta_1) \hat{W}(t_2, \Psi_2, \Phi_2, \Theta_2) \hat{V}(t_1, \Psi_1, \Phi_1, \Theta_1)  \rangle, 
    \label{eq:OTOC}
\end{equation}
where capital letters denote boundary coordinates, $t_2 - t_1 \gg \beta$ and the angled brackets denote the trace with respect to the thermal density matrix $e^{-\beta H}/Z$ in the dual field theory.  As is now well known, this OTOC can be computed by considering the gravitational scattering amplitude between a particle of momentum $p_1^{U}$ travelling along the $V=0$ horizon of the black hole (corresponding to quanta created by the $\hat{V}$ operator) and a particle of momentum $p_2^{V}$ travelling along the $U=0$ horizon (corresponding to quanta created by the $\hat{W}$ operator). At the level of the eikonal approximation, the scattering amplitude for this process takes the form $e^{i\delta}$ where the eikonal phase $\delta$ is given by evaluating the Einstein-Hilbert action on gravitational shockwave solutions sourced by the $\hat{V}$ and $\hat{W}$ operators \cite{shenker2015stringy}. The OTOC is then given by an integral, over momenta and angular coordinates, of $e^{i\delta}$ weighted by bulk-boundary wave functions, which describe the distribution of quanta along at the horizon.

\paragraph{} We begin by considering the quanta created by the $\hat{V}$ operator, corresponding to a particle of momentum $p_1^{U}$, whose trajectory approximates the null geodesic given by $(U = U(\tau), V=0,  \tilde{\psi}_1, \phi_1, \theta_1)$\footnote{Note the fact we are in co-rotating coordinates is important for this to be a geodesic.}. The only non-zero component of the stress tensor of such a particle is given by:

\begin{equation}
    T^{UU} = \frac{1}{\sqrt{-g}}p_1^U \delta(V)\delta(\tilde{\psi} - \tilde{\psi}_1)\delta(\phi - \phi_1)\delta(\theta - \theta_1),
    \label{V-quant}
\end{equation}
Likewise, the quanta created by the $\hat{W}$ operator can be represented by a second particle of momentum $p_1^{V}$ following the null-geodesic $(U=0, V = V(\tau), \tilde{\psi}_2, \phi_2, \theta_2)$, with the stress tensor
\begin{equation}
    T^{VV} = \frac{1}{\sqrt{-g}}p_2^V \delta(U)\delta(\tilde{\psi}-\tilde{\psi}_2)\delta(\phi - \phi_2)\delta(\theta - \theta_2).  
    \label{W-quant}
\end{equation}
The eikonal phase $\delta$ is then given by determining the linearized gravitational backreaction $\delta g_{\mu \nu}$ sourced by \eqref{V-quant} and \eqref{W-quant}, and evaluating the quadratic Einstein Hilbert action on these solutions \cite{shenker2015stringy}:
\begin{equation}
    \delta = \frac{1}{4} \int d^5 x \sqrt{-g} \delta g_{\mu\nu} T^{\mu \nu} = \frac{1}{4} \int d^5 x \sqrt{-g} (\delta g_{UU} T^{UU} + \delta g_{VV}T^{VV}). 
    \label{delta}
\end{equation}
\paragraph{} We begin by computing the gravitational back-reaction due to \eqref{V-quant}.  After lowering indices with \eqref{Kruskal} the only non-zero component of the stress tensor sourced by the $\hat{V}$ quanta is $T_{VV} = A(0)^2 T^{UU}/4$.  As usual the gravitational backreaction of this source can be computed by considering an ansatz for a geometry corresponding to a shockwave across the $U = 0$ horizon
\begin{equation}
    \delta g_{VV} = -A(0) f_1(\tilde{\psi}, \phi, \theta)\delta (V). 
    \label{Shock-ansatz}
\end{equation}
Inserting the shockwave ansatz (\ref{Shock-ansatz}) into Einstein's equations one finds the only non-trivial component of Einstein's equations is the $VV$-component, which reduces to the following partial differential equation for the angular profile $f_1(\tilde{\psi}, \phi, \theta)$:
\begin{equation}
   \mathcal{D} f_1(\tilde{\psi}, \phi, \theta ) = \frac{16 \pi G_N}{\mathrm{sin}(\theta_1)}\frac{K}{L r_0^2} p_1^U \delta(\tilde{\psi} - \tilde{\psi}_1)\delta(\phi - \phi_1)\delta(\theta - \theta_1), 
   \label{shockwave}
\end{equation}
where the differential operator $\mathcal{D}$ is given by:
\begin{equation}
\square  + \lambda_1 + \lambda_2 \partial_{\tilde{\psi}} + \lambda_3 \partial^2_{\tilde{\psi}}, 
     \label{shock-diff}
\end{equation}
with $\square$ one quarter of the Laplacian on the unit 3-sphere written in terms of the Hopf fibration  
\begin{equation}
    \square =  \mathrm{cot}(\theta)\partial_{\theta} +  \partial_{\theta}^2 + \mathrm{csc}^2(\theta) \partial_{\tilde{\psi}}^2 +\mathrm{csc}^2(\theta)\partial_{\phi}^2 -2\mathrm{cot}(\theta)\mathrm{csc}(\theta)\partial_{\tilde{\psi}}\partial_{\phi},
\end{equation}
and $\lambda_1, \lambda_2, \lambda_3$ constants given by the expressions
\begin{equation}
    \lambda_1 =  \frac{(2K^2 + L^2 r_0^2)(2a^2(r_0^2 + L^2)^2 -\Delta)}{4L^4r_0^2 K^2}, \hspace{0.5cm} \lambda_2 = -\frac{2 a (r_0^2 + L^2) K}{L^3 r_0^2}, \hspace{0.5cm} \lambda_3 = -\frac{a^2(r_0^2 + L^2)}{L^2 r_0^2}.
    \label{lambdas}
\end{equation}

The $W$ quanta travelling on the $U = 0$ horizon source an analogous metric perturbation. The calculation mirrors the one performed above, with the shockwave giving rise to a Kruskal geometry with a shift across the $U=0$ horizon $V \to V + f_2(\tilde{\psi}, \phi, \theta)$. To linear order in $f_2$ this leads to the following metric perturbation
\begin{equation}
\delta g_{UU} = -A(0)\delta(U)f_2(\tilde{\psi}, \phi, \theta),
\end{equation}
which solves Einstein's equations sourced by (\ref{W-quant}) subject to $f_2(\tilde{\psi}, \phi, \theta)$ satisfying the following equation
\begin{equation}
    \tilde{\mathcal{D}} f_2(\tilde{\psi}, \phi, \theta) =\frac{16 \pi G_N}{\mathrm{sin}(\theta_2)} \frac{K}{L r_0^2} p_2^V \delta(\tilde{\psi} - \tilde{\psi}_2)\delta(\phi - \phi_2)\delta(\theta - \theta_2),
   \label{shock-diff2} 
\end{equation}
where $\tilde{\mathcal{D}}$ is now the differential operator given in (\ref{shock-diff}) but with $\partial_{\tilde{\psi}} \to -\partial_{\tilde{\psi}}$ and $\partial_\phi \to -\partial_\phi$.\\

We are now in a position to compute the eikonal phase $\delta$ using equation (\ref{delta}). To do so, we first introduce the function $f(\tilde{\psi}, \phi, \theta, \theta')$ as the Green's function of the Laplacian on $S^3$ with normalised delta-function source
\begin{equation}
\mathcal{D} f(\tilde{\psi}, \phi, \theta, \theta') = -\frac{1}{2 \sin(\theta')} \delta(\tilde{\psi}) \delta(\phi) \delta(\theta - \theta')
\label{normalisedshock}
\end{equation}
in terms of which we find that the action \eqref{delta} evaluated on \eqref{shock-diff} and \eqref{shock-diff2} gives rise to the eikonal phase\footnote{Note here we have used translational symmetry in the $\phi, \tilde{\psi}$ directions and that $\tilde{f}(\tilde{\psi}_1 - \tilde{\psi}_2, \phi_1 - \phi_2, \theta_1, \theta_2) = f(\tilde{\psi}_2 - \tilde{\psi}_1, \phi_2 - \phi_1, \theta_2, \theta_1)$ with $\tilde{f}$ the solution to the analogous equation to \eqref{normalisedshock} with $\mathcal{D}  \to \tilde{\mathcal{D}}$.} 
 \begin{equation}
     \delta = \frac{16 \pi G_N  K}{L r_0^2} A(0) p_1^{U} p_2^{V} f(\psi_2 - \psi_1, \phi_2 - \phi_1, \theta_2, \theta_1) 
 \end{equation}
Due to the large time separation the between the $\hat{V}, \hat{W}$ the typical centre of mass of energy of the collision grows exponentially in time as $p_1^{U} p_2^{V} \sim e^{2 \pi T t}$. In co-rotating coordinates the eikonal phase therefore grows exponentially in time with an exponent $2 \pi T$ and has an angular profile governed by the solution $f(\tilde{\psi}, \phi, \theta, \theta')$ to the shockwave equation. 

\subsection*{Exact solution for angular profile of shockwave}
It is possible to obtain an exact analytic solution to \eqref{normalisedshock} as an expansion in Wigner D-functions. These are the eigenfunctions of the Laplacian on $S^3$, and provide a complete set of functions analogous to the case of spherical harmonics for the 2-sphere. In particular, the Wigner D-functions are indexed by either integer or half-integer ${\cal J}$, together with quantum numbers ${\cal K}, {\cal M}$ such that $|{\cal M}| \leq {\cal J}, |{\cal K}| \leq {\cal J}$. The Wigner D-functions then have the form $D^{\cal J}_{{\cal K}{\cal M}}(\tilde{\psi}, \phi,  \theta) = d^{\cal J}_{{\cal K} {\cal M}}(\theta) e^{i {\cal K} \tilde{\psi} + i \cal{M} \tilde{\phi}}$ and satisfy
\begin{equation}
\square D^{\cal J}_{{\cal K}{\cal M}} + {\cal J}({\cal J }+ 1)  D^{\cal J}_{{\cal K}{\cal M}} = 0. 
\label{wignerdeqn}
\end{equation}
They form a complete set of functions on the three-sphere, with the completeness relation\cite{Khersonskii:1988krb}
\begin{equation}
\frac{1}{\sin(\theta_1)} \delta(\tilde{\psi} - \tilde{\psi}_1) \delta(\phi - \phi_1) \delta(\theta - \theta_1) = \sum_{{\cal J}=0, 1/2, 1, \dots}^{\infty} \sum_{{\cal K}=-{\cal J}}^{\cal J}  \sum_{{\cal M}={-{\cal J}}}^{{\cal J}} \frac{2 {\cal J}+1}{16 \pi^2} D^{{\cal J}}_{{\cal K} {\cal M}}(\tilde{\psi}, \phi, \theta) D^{{\cal J}*}_{{\cal K} {\cal M}}(\tilde{\psi_1}, \phi_1, \theta_1) \nonumber.
\label{completeness}
\end{equation}
Using this completeness relation to expand the delta function in \eqref{normalisedshock} in terms of Wigner D-functions, and making use of \eqref{wignerdeqn}, we find that the normalised shockwave profile is given by
\begin{equation}
f(\tilde{\psi}, \phi, \theta, \theta^{\prime}) =  \sum_{{\cal J}=0,1/2,1,\dots}^{\infty} \sum_{{\cal K}=-{\cal J}}^{{\cal J}}  \sum_{{\cal M}=-{\cal J}}^{{\cal J}}     \frac{2{\cal J}+1}{16  \pi^2}  \frac{d^{{\cal J}}_{{\cal K} {\cal M}}(\theta^{\prime})  d^{{\cal J}}_{{\cal K} {\cal M}}(\theta)}{{\cal J}({\cal J}+1)- \lambda_1 - i \lambda_2 {\cal K} + \lambda_3 {\cal K}^2}  e^{i {\cal K} \tilde{\psi} + i {\cal M} \phi}.  
\label{exactshock}
\end{equation}
We therefore obtain an exact expression for the shock-wave equation in the Myers-Perry-AdS$_5$ black hole as a sum over Wigner D-functions. In the next section we will see that in case of large black holes it is possible to perform this infinite sum analytically and obtain an explicit expression for the shockwave profile and hence OTOC for operator insertions lying on Hopf circles of the three-sphere, which corresponds to setting $\theta = \theta' = 0$ in \eqref{exactshock}. 
 
 \subsection{OTOC on Hopf circles}\label{Sect_OTOC_Hopf}
 
 In the last subsection we presented an exact expression~\eqref{exactshock} for the shock-wave profile governing the form of the eikonal phase related to the OTOC. For general configurations $(\psi, \phi, \theta, \theta^{\prime})$ we are not aware of a simple way of computing the sum in \eqref{exactshock}. However, a particularly interesting configuration of the OTOC that we can directly analyse are when external operators $\hat{V}$ and $\hat{W}$ lie on a Hopf circle, that is they have the same coordinates on the base $S^2$ and are separated only in the direction of the fibre coordinate $\psi$ that is parallel to the rotation.  Concretely, we therefore consider OTOCs of the form:
 \begin{equation}
     H(t, \Psi) = \langle W(t_2, \Psi_2, \Theta, \Phi)V(t_1, \Psi_1, \Theta, \Phi)W(t_2, \Psi_2, \Theta, \Phi)V(t_1, \Psi_1, \Theta, \Phi)\rangle
 \end{equation}
where $t = t_2 - t_1$ and $\Psi = \Psi_2 - \Psi_1$. Such operator configurations can be mapped by an isometry  to the North pole, and henceforth we therefore set $\Theta = \Phi = 0$ without loss of generality.  

\paragraph{} For such operator configurations, then in the large black hole limit it is possible to obtain a simple closed form expression for the OTOC. In particular, in the large black hole limit $r_0 \gg L$ the bulk to boundary wave-functions describing the distribution of quanta on the horizon will be sharply peaked in the angular coordinates on the horizon around $\tilde{\psi}_2 - \tilde{\psi}_1 = \Psi - \Omega t$, $\theta_1 \approx \theta_2 \approx 0$, $\phi_1 \approx \phi_2 \approx 0$. Expanding the standard expression from \cite{shenker2015stringy} for the OTOC in terms of the eikonal phase to leading order in $G_N$, and using that the bulk-to-boundary wave-functions are sharply peaked around $p_1^{U} p_2^{V} \sim e^{2 \pi T t}$, the OTOC then takes the form 
\begin{equation}
\label{otocsimple}
    H(t, \Psi) \sim 1 - c G_N e^{2\pi T t}f(\Psi - \Omega t).
\end{equation}
Here $c$ is a constant, $f(\tilde{\psi})$ is the solution to equation (\ref{normalisedshock}) with $\theta  = \theta' = \phi = 0$ and we note that in the large black hole limit the expressions~\eqref{temperature} and \eqref{omega} reduce to
\begin{equation}
2 \pi T = \frac{2 r_0}{L^2} \sqrt{1 - a^2/L^2}, \hspace{1.0cm} \Omega = -2 a/L^2.
\end{equation}
As we show in Appendix~\ref{appendix:Solve_shock}, in the large black hole limit the expression \eqref{exactshock} can be approximated by an integral, leading to the expression 
\begin{equation}
f(\tilde{\psi}) =  \int_{0}^{\infty} d {\cal J} \int_{-{\cal J}}^{{\cal J}} d {\cal K}   \frac{{\cal J}}{4 \pi^2}  \frac{e^{i {\cal K} \tilde{\psi}}}{{\cal J}^2- \lambda_1 - i  \lambda_2 {\cal K} + \lambda_3 {\cal K}^2}
\end{equation}
for the shock-wave profile on Hopf circles. This integral can be computed exactly by contour integration (see Appendix \ref{appendix:Solve_shock} for details), to give 
\begin{eqnarray}
    f(\tilde{\psi}) &=& -\frac{e^{-k_+ \tilde{\psi}}}{4\pi |\tilde{\psi}|}, \hspace{1.0cm} \tilde{\psi} > 0, \nonumber \\
 f(\tilde{\psi})  &=&    -\frac{e^{k_- \tilde{\psi}}}{4\pi  |\tilde{\psi}|}, \hspace{1.0cm} \tilde{\psi} < 0,
    \label{hopfotocF}
\end{eqnarray}
where 
\begin{equation}
    k_{\pm} =  \frac{r_0}{L\sqrt{1 - a^2/L^2}}(\sqrt{\frac{3}{2}} \mp \frac{a}{L}). 
\end{equation}
In co-rotating coordinates, the OTOC is therefore exponentially growing in time with an exponent $2 \pi T t$ and exponentially decaying in the $\tilde{\psi}$ direction governed by $k_{\pm}$. In terms of the fixed boundary coordinates $(t, \Psi)$ the exponentially growing piece of the OTOC~\eqref{otocsimple} takes the functional form 
\begin{eqnarray}
    H(t, \Psi) &\sim& -\frac{\exp{\bigg(2 \pi T_+\bigg(t - \frac{L \Psi}{2 v_B^{+}}\bigg)}\bigg)}{4\pi |\Psi - \Omega t|}, \hspace{1.0cm} \Psi - \Omega t > 0, \nonumber \\
 H(t, \Psi)  &\sim&   -\frac{\exp{\bigg(2 \pi T_-\bigg(t - \frac{L \Psi}{2 v_B^{-}}\bigg)}\bigg)}{4\pi |\Psi - \Omega t|}, \hspace{1.0cm} \Psi - \Omega t < 0,
    \label{hopfotoc}
\end{eqnarray}
where in the large black hole limit $\Omega = - 2 a/L^2$ , $L \Psi/2$ corresponds to the spatial distance between the operators $\hat{V}, \hat{W}$\footnote{Note $\psi$ runs from $0$ to $4\pi$ over a circle of radius $L$.} and the exponents $T_{\pm}, v_B^{\pm}$ are given by 
\begin{equation}
2 \pi T_{\pm} = \frac{2 r_0}{L^2 \sqrt{1 - a^2/L^2}} (1 \mp \sqrt{\frac{3}{2}} \frac{a}{L}),
\label{rotatinglyapunov}
\end{equation}
and 
\begin{equation}
\frac{2 \pi T_{\pm}}{v_B^{\pm}} =  \frac{2 r_0}{L^2\sqrt{1 - a^2/L^2}}(\sqrt{\frac{3}{2}} \mp \frac{a}{L}). 
\label{rotatingvb}
\end{equation}
The form of the above expressions can be understood by applying a boost of velocity $v = -\Omega L/2 = a/L$ to the expression $H_4 \sim \exp(2 \pi T_0(t - L \Psi/2 v_B^{(0)}))$ for the OTOC of the AdS$_5$-Schwarzschild black brane. Under such a boost one finds 
\begin{equation}
2 \pi T_{\pm} = 2 \pi T_0 \gamma_v\bigg(1 \mp \frac{v}{v_B^{(0)}}\bigg), \hspace{1.0cm} \frac{2 \pi T_{\pm}}{v_B^{\pm}} = 2 \pi T_0 \gamma_v\bigg(\frac{1}{v_B^{(0)}} \mp v \bigg), 
\end{equation}
which are equivalent to \eqref{rotatinglyapunov} and \eqref{rotatingvb} upon using the expressions $2 \pi T_0 = 2 r_0/L^2$ and $v_B^{(0)} = \sqrt{2/3}$, with $v_B^{(0)}$ the butterfly velocity of the static Schwarzschild-AdS$_5$ black brane (which we refer to as the conformal butterfly velocity). 

 \section{Near horizon ingoing metric perturbations}\label{NH}
\paragraph{} We now wish to study the energy density response of the field theory dual to the geometry given by (\ref{metric}). In order to do this, we will study in-going metric perturbations about (\ref{metric}) from which we can extract information about the retarded Green's function $G^{R}_{T^{00}T^{00}}$. We will see that this exhibits the characteristic features of pole-skipping as identified in \cite{2018G, 2018B, 2018}. In particular, we will show that at a special value of the frequency $\omega = i2\pi T$ the linearized Einstein equations admit an additional ingoing mode when the angular profile of the metric component $\delta g_{vv}$ at the horizon is solution to the (sourceless) shockwave equation \eqref{shockwave} that governs the form of the OTOC. 

\paragraph{} In particular, we work in ingoing coordinates $(v, r, \tilde{\psi}, \phi, \theta)$ where $v$ and $\tilde{\psi}$ are defined in (\ref{ingoing}) and (\ref{co-rot}) respectively. We then study linearized perturbations of the metric (\ref{metric}), and for convenience perform a Fourier transform with respect to $v$ and the co-rotating angle $\tilde{\psi}$:
\begin{equation}
\label{metpert}
\delta g_{\mu \nu}(v, r, \tilde{\psi}, \phi, \theta ) = e^{i (k \tilde{\psi} -\omega v)}\delta g_{\mu \nu}(r, \phi, \theta)
\end{equation}
Inserting this ansatz into the linearized Einstein equations will give rise a coupled set of coupled PDEs for $\delta g_{\mu \nu}(r, \phi, \theta)$. However, a significant simplification of the equations is made possible dual to the enhanced symmetry of the simply-spinning black hole. In particular, after parameterising the spatial profile of the metric components in terms of Wigner D-functions the Einstein equations can be reduced to a coupled set of ODEs for the radial profiles of each metric component~\cite{Murata:2007gv,Murata:2008yx,Garbiso:2020puw}, which we discuss further in Section~\ref{Num}. 

\paragraph{} For the purposes of identifying pole-skipping however, it is sufficient to work directly with the ansatz~\eqref{metpert} and analyse the near horizon behaviour of the linearized Einstein equations. In particular, as in previous examples of pole-skipping, we find that at $\omega = i2\pi T$ the $vv$ component of the Einstein equations at $r = r_0$ reduces to a decoupled equation for the angular profile of $\delta g_{vv}$ at the horizon. In particular, all other metric perturbation decouple from this component of the Einstein equations, which evaluates at the horizon to
\begin{equation}
       ( (\mathrm{cot}(\theta)\partial_{\theta} +  \partial_{\theta}^2 - \mathrm{csc}^2(\theta) k^2 + \mathrm{csc}^2(\theta)\partial_{\phi}^2 -2i\mathrm{cot}(\theta)\mathrm{csc}(\theta)k\partial_{\phi} ) + \lambda_1 + i\lambda_2 k - \lambda_3 k^2) \delta g_{vv}(r_0, \phi, \theta) = 0
       \label{horEqn}
\end{equation}
where $\lambda_1$, $\lambda_2$ and $\lambda_3$ the same constants defined in equation (\ref{lambdas}). This equation \eqref{horEqn} is precisely the source-less version of the shockwave equation governing the form of OTOC (\ref{shockwave}), after a Fourier transform with respect to the $\psi$ coordinate. The key point is that the metric perturbation $\delta g_{vv}(r_0, \phi, \theta)$ must either have a specific angular profile on the horizon (i.e. be a non-trivial solution to \eqref{horEqn}) or it must be zero at the horizon. As such, for metric perturbations with a particular angular profile related to the shockwave equation we expect an extra ingoing mode, and the pole-skipping phenomenon. 

\subsection{Pole-skipping criterion in terms of Wigner D-functions}

As we mentioned in the previous subsection, in the simply spinning Myers-Perry AdS black hole the angular and radial dependence of the linearized Einstein equations can be decoupled by considering metric perturbations whose angular profile is governed by a Wigner D-function~\cite{Murata:2007gv,Murata:2008yx,Garbiso:2020puw}. It is instructive to ask how these Wigner D-functions are related to the angular profiles satisfying \eqref{horEqn} for which we expect pole-skipping.

\paragraph{} In particular, let us recall that the Wigner D-functions are the eigenfunctions of the differential operator $\square$. I.e. we have  $\mathcal{D}^{{\cal J}}_{{\cal K} {\cal M}}(\tilde{\psi}, \theta, \phi) = d^{{\cal J}}_{{\cal K} {\cal M}}(\theta) e^{i {\cal K} \tilde{\psi} + i {\cal M \phi}}$ where 
\begin{equation}
\label{D-function}
\square \mathcal{D}^{{\cal J}}_{{\cal K} {\cal M}} + {\cal J}({\cal J}+1) \mathcal{D}^{\cal J}_{{\cal K}{\cal M} } = 0. 
\end{equation} 
For integer and half-integer ${\cal J}$ with $|{\cal M}|, |{\cal K}| \leq {\cal J}$ there is a unique solution to \eqref{D-function} that is regular on $S^3$. In order to consider pole-skipping, it is necessary to relax the condition that ${\cal M}, {\cal K}, {\cal J}$ be integers and consider them to be arbitrary complex parameters. This corresponds to considering angular profiles of the metric components that are not necessarily regular on the three-sphere. 
\paragraph{} In particular, the horizon equation (\ref{horEqn}) is equivalent to the statement that at pole-skipping points the angular profile of $\delta g_{vv}$ satisfies \eqref{D-function} but where the quantum numbers ${\cal K}$ and ${\cal J}$ are related by
\begin{equation}
\label{poleskippingW}
\lambda_1 + i \lambda_2 {\cal K} - \lambda_3 {\cal K}^2 = {\cal J}({\cal J} + 1). 
\end{equation}
Note that, as a result of the equivalence between the horizon equation and shockwave equation, this relationship corresponds exactly to the locations at which the summand in the shockwave profile \eqref{exactshock} has poles. In the large black hole limit, we can use this to obtain a precise connection between the locations of pole-skipping points and the form of the OTOC on Hopf circles computed in \eqref{hopfotoc}. 
\paragraph{} In order to compare to \eqref{hopfotoc} we again take the large black hole limit $r_0/L \gg 1$ while keeping $a/L \sim \mathcal{O}(1)$. In the this limit the constants $\lambda_1$, $\lambda_2$, $\lambda_3$ simplify to:
\begin{equation}
\lambda_1 = \frac{r_0^2}{2L^4}(2a^2 -3L^2), \hspace{1cm} \lambda_2 = -\frac{2ar_0}{L^2} \sqrt{1-\frac{a^2}{L^2}} \hspace{1cm} \lambda_3 = -\frac{a^2}{L^2}. \nonumber
\end{equation}
We now consider the class of pole-skipping points where the quantum number ${\cal K}$ parallel to the rotation direction takes its maximal value, i.e. we set ${\cal K} = {\cal J}$. Further, using that in the large black hole limit we have ${\cal K} \sim r_0/L \gg 1$, we find that \eqref{poleskippingW} reduces to the following quadratic equation for ${\cal K}$
\begin{equation}
(1 - \frac{a^2}{L^2} ){\cal K}^2 + \frac{2 i a r_0}{L^2} \sqrt{1 -\frac{a^2}{L^2}} {\cal K} + \frac{r_0^2}{2L^4}(3 L^2 - 2a^2) = 0
\end{equation}
whose solutions are given by:
\begin{equation}
\label{wavevectors}
{\cal K} = \pm \frac{i r_0}{L\sqrt{1 - a^2/L^2}}\bigg(\sqrt{\frac{3}{2}} \mp \frac{a}{L} \bigg) = \pm i k_{\pm}. 
\end{equation}
We therefore see these modes exhibit pole-skipping at a frequency $\omega = i 2 \pi T$ and at precisely the same wave-vectors $k_{\pm}$ that appear in the exponential form of the OTOC on Hopf circles computed in \eqref{hopfotoc}.

\paragraph{} Finally, we note that so far we have been working in terms of co-rotating coordinates, which corresponds to studying the boundary response in coordinates $(t, \tilde{\psi}, \theta, \phi)$. If we transform these results to the standard coordinates $(t, \psi, \theta, \phi)$ for the three sphere then we find that the wave-vectors $k_{\pm}$ at which pole-skipping occurs are unchanged, whilst the frequencies at which pole-skipping occurs are modified to 
\begin{equation} \label{eq:poleSkippingPointsRestFrame}
\omega_{\pm} = i(2\pi T \pm k_{\pm}\Omega) = \frac{2 ir_0}{L^2\sqrt{1 - a^2/L^2}}\bigg(1 \mp \sqrt{\frac{3}{2}} \frac{a}{L}\bigg). 
\end{equation}
As we will shortly discuss in more detail, the locations $(\omega, {\cal K}) = (\omega_\pm, \pm i k_\pm)$ at which pole-skipping occurs in these coordinates are precisely what one obtains by applying a boost to the pole-skipping points of the AdS$_5$ black brane, and correspond precisely to the frequencies and wave-vectors that appear in the profile of the OTOC on Hopf-circles that we computed in Section~\ref{Sect_OTOC_Hopf}.

\section{Full perturbation equations and numerical checks of pole-skipping}\label{Num}

\paragraph{} In the last Section we used a near-horizon analysis to argue that for the existence of pole-skipping in the retarded energy density Green's function when the quantum numbers ${\cal J}, {\cal K}, {\cal M}$ parameterising the angular profile of fluctuations satisfy \eqref{poleskippingW}. Here we will numerically confirm our results for the large black hole limit, in which we argued that pole skipping locations with ${\cal K} = {\cal J}$ are given in the rest frame of the boundary theory by $(\omega, {\cal K}) = (\omega_\pm, \pm i k_\pm)$ defined through \eqref{wavevectors} and \eqref{eq:poleSkippingPointsRestFrame}. 

\paragraph{} To do this, we review the structure of the gravitational perturbation equations around the rotating black hole~\eqref{metric} in more detail. In particular, as was first discussed in~\cite{Murata:2007gv,Murata:2008yx}, general bulk metric perturbations $\delta g_{\mu \nu}$ can be parameterised in terms of the Wigner-D functions and radial functions $\delta h_{a b}(r)$ where $a,b$ run over $t,r$ and the indices $+, -, 3$ associated with the one forms in \eqref{metric}. For given quantum numbers one then obtains a set of coupled ODEs for $\delta h_{a b}(r)$. Solving these ODEs subject to ingoing boundary conditions then allows one to extract the retarded Green's functions of the boundary stress tensor.

\paragraph{} A significant simplification in the perturbation equations occurs if we restrict the parameters in the Wigner functions as $\mathcal{K} = \mathcal{J}$, which can be interpreted as choosing the angular momentum of the fluctuations to be parallel to angular momentum of the background rotation. Upon doing so one finds that the system of fluctuations equations for $\delta h_{a b}(r)$ splits into decoupled scalar, vector, and tensor sectors~\cite{Murata:2008xr} due to the aforementioned $SU(2)\times U(1)$ symmetry of the rotating black hole metric~\eqref{metric}.

\paragraph{} Furthermore, we now consider the large black hole limit defined by taking~\cite{Garbiso:2020puw,Cartwright:2021qpp} 
\begin{equation}
    (r_0/L) \gg 1 \, : \quad \
\omega\to (r_0/L) \, 2 \nu \, , \quad \
\mathcal K\to (r_0/L)\, \mathcal K \, , \quad \
\mathcal J\to (r_0/L)\, j \, , \label{eq:largeBHLimit}
\end{equation}
simultaneously 
with the black hole horizon radius and radial coordinate scaling given by 
\begin{equation}\label{eq:largeBHLimitCoor}
    (r_0/L)\to \infty \, , \quad 
    r\to (r_0/L)\,  r\, , 
\end{equation}
and keep only leading order terms in $(r_0/L)$. After these steps, the system of perturbation equations of the three channels can be reduced to three decoupled ordinary differential equations, each for one gauge invariant master field~\cite{Garbiso:2020puw,Cartwright:2021qpp}. 

\paragraph{} In the scalar channel, this yields the fluctuation equation (after setting $L=1$ for simplicity of presentation in the remainder of this section)
\begin{equation}\label{eq:scalar_sector}
    \begin{aligned}
       0 &= \frac{\left(a^2 \left(4 u^2-u \nu^2-4\right)-2 a j u \nu-u \left(j^2+4 u\right)+4\right)}{a^2-1}Z_0(u) + \\
         &\quad\quad\frac{f(u)}{u}\left(3 u^2-5\right) Z_0'(u) +\frac{f(u)^2}{u^2} Z_0''(u) ,   
    \end{aligned}
\end{equation}
where $Z_0(u)$ is the master variable formed from the coupled fluctuations
\begin{align}
     Z_0(u) &= (2  \left(j^2 \left(a^2-z^4-1\right)-a^2 \nu ^2 \left(z^4+1\right)-2 a j \nu  z^4+\nu ^2\right) h_{+-}(u)\nonumber\\
        &+\left(a^2-1\right)  (j^2 h_{tt}(u) - 4j \nu  h_{t3}(u) +4\nu^2 h_{33}(u)))
\end{align}
where the $\pm$-coordinate-directions are defined by $\sigma_\pm = \frac{1}{2}(\sigma_1\mp i \sigma_2)$, 
and we recall that the parameter, $a$, quantifies the rotation of the background metric.\footnote{While the $\sigma_3$-direction is associated with eigenvalues $\mathcal{K}$ of the $SU(2)\times U(1)$-operator $W_3$ through $W_3 D^{\mathcal{J}}_\mathcal{K M} = \mathcal{K} D^{\mathcal{J}}_\mathcal{K M}$, the $\sigma_\pm$-directions are associated with ladder operators $W_{\pm}=W_1\pm i W_2$ of that $SU(2)\times U(1)$-algebra~\cite{Murata:2008xr,Murata:2008yx,Garbiso:2020puw}.}  

\paragraph{} Now we note  that the explicit dependence on the rotation parameter $a$ can be removed with a linear redefinition of the frequency and momentum
\begin{equation}\label{eq:criticalPointTrajectory}
    \begin{aligned}
         \mathfrak{q}^2 &= \frac{\left(a \nu + j \right)^2}{1-a^2}\,, \qquad \mathfrak{q} = \frac k {2 \pi T},\,\\
         \mathfrak{w}^2 &= \frac{\left(\nu + a j \right)^2}{1-a^2}\,.  \qquad
         \mathfrak{w} = \frac \omega {2 \pi T}\ \, ,  
    \end{aligned}
\end{equation}
after which the resulting $a$-independent equation is equivalent to the equation for the master field governing (scalar sector, spin-0) metric perturbations around a Schwarzschild AdS black brane~\cite{Kovtun:2005ev}. 
Furthermore, the linear definition~\eqref{eq:criticalPointTrajectory} is simply a boost transformation of the frequency $\nu$ and wavevector $j$ of the fluctuations. 
 
\paragraph{} Hence, \eqref{eq:scalar_sector} is simply the master equation for scalar fluctuations around a boosted black brane, where the momentum of the fluctuations is aligned with the boost direction. 
In fact, in all three sectors (scalar, vector, tensor) of the rotating system then in the large black hole limit metric perturbations with ${\cal K} = {\cal J}$ can be reduced in the same way to fluctuations around a boosted Schwarzschild AdS black brane. This dramatically simplifies the calculation of the QNMs. In particular, we can directly compute the pole-skipping locations simply by boosting the known pole-skipping locations of the Schwarzschild AdS black brane.  

\paragraph{} Note, however, that there is only one of the four solutions to equation~\eqref{eq:criticalPointTrajectory} 
which correctly reduces to the statement that $\mathfrak{q}=j$ and $\mathfrak{w}=\nu$ in the limit that $a \rightarrow 0$. 
This branch is the physical branch, given by
\begin{equation} \label{eq:physicalBoost}
    \mathfrak{q}= \dfrac{a \nu +j}{\sqrt{1-a^2}},\quad \mathfrak{w}= \dfrac{a j+\nu }{\sqrt{1-a^2}} \, .
\end{equation}
This is the solution we choose to transform the numerically known pole-skipping points of the AdS Schwarzschild black brane~\cite{Grozdanov_2019} to the pole-skipping points of the rotating black hole in the large black hole limit~\eqref{eq:largeBHLimit}. 

\subsubsection*{Pole-skipping locations}

\paragraph{} As discussed above, the locations of the pole-skipping points for a system at rest can be continued into the rotating pole-skipping points, in the large black hole limit~\eqref{eq:largeBHLimit}, using the relation given in \eqref{eq:physicalBoost}. Here we will discuss the pole-skipping locations in the scalar channel, which are relevant for a comparison with the results of Section~\ref{NH} and are those expected to be directly related to the OTOC. However one also perform analogous calculations to analyse other pole-skipping locations in the vector and tensor channel, and we briefly present such results in Appendix~\ref{appendix:otherchannels}. 

\paragraph{} In particular, for a non-rotating Schwarzschild AdS black brane geometry the pole-skipping points in the scalar channel are given by~\cite{shenker2015stringy}
\begin{equation}
    \mathfrak{w} = i,\quad \mathfrak{q}= \pm  \sqrt{\frac{3}{2}} i\, .
\end{equation}
Utilizing equation~(\ref{eq:physicalBoost}) this gives that the pole-skipping points for the rotating system in the large black hole limit are 
\begin{equation}\label{eq:Scalar_Pole_Skipping}
  \nu_{scalar}= \frac{i}{\sqrt{1-a^2/L^2}}\left(1\mp\frac{\sqrt{3}}{\sqrt{2}} \frac{a}{L}\right),\quad j_{scalar}= \frac{i}{\sqrt{1-a^2/L^2}}\left(\pm\frac{\sqrt{3}}{\sqrt{2}}-  \frac{a}{L}\right)
  \, , 
\end{equation}
where we have restored factors of $L$ for ease of comparison with the pole-skipping points in the rest frame given in Section~\ref{NH}. Using $\mathfrak{w} = \omega/2\pi T, \mathfrak{q} = k/2\pi T$ one sees these results are identical to those for $(\omega, {\cal K}) = (\omega_\pm, \pm i k_\pm)$ in equation~\eqref{eq:poleSkippingPointsRestFrame}, confirming, as we previously remarked, that the pole-skipping locations are simply those of a boosted black brane. 

\paragraph{} As well as boosting the pole-skipping points of the Schwarzschild AdS black brane, we have verified explicitly verified pole-skipping at these locations by numerically solving for the dispersion relation of the (sound) QNM in the scalar sector of perturbations for each value of $a/L$.  We have done this using two independent methods from two independently constructed routines, firstly from a static Schwarzschild black brane whose modes are then boosted and secondly directly from the large black hole limit of the rotating black hole (for further details see Appendix~\ref{appendix:numericalConsistency}).

\begin{figure}[htbp]
    \centering
    \includegraphics[width=12cm]{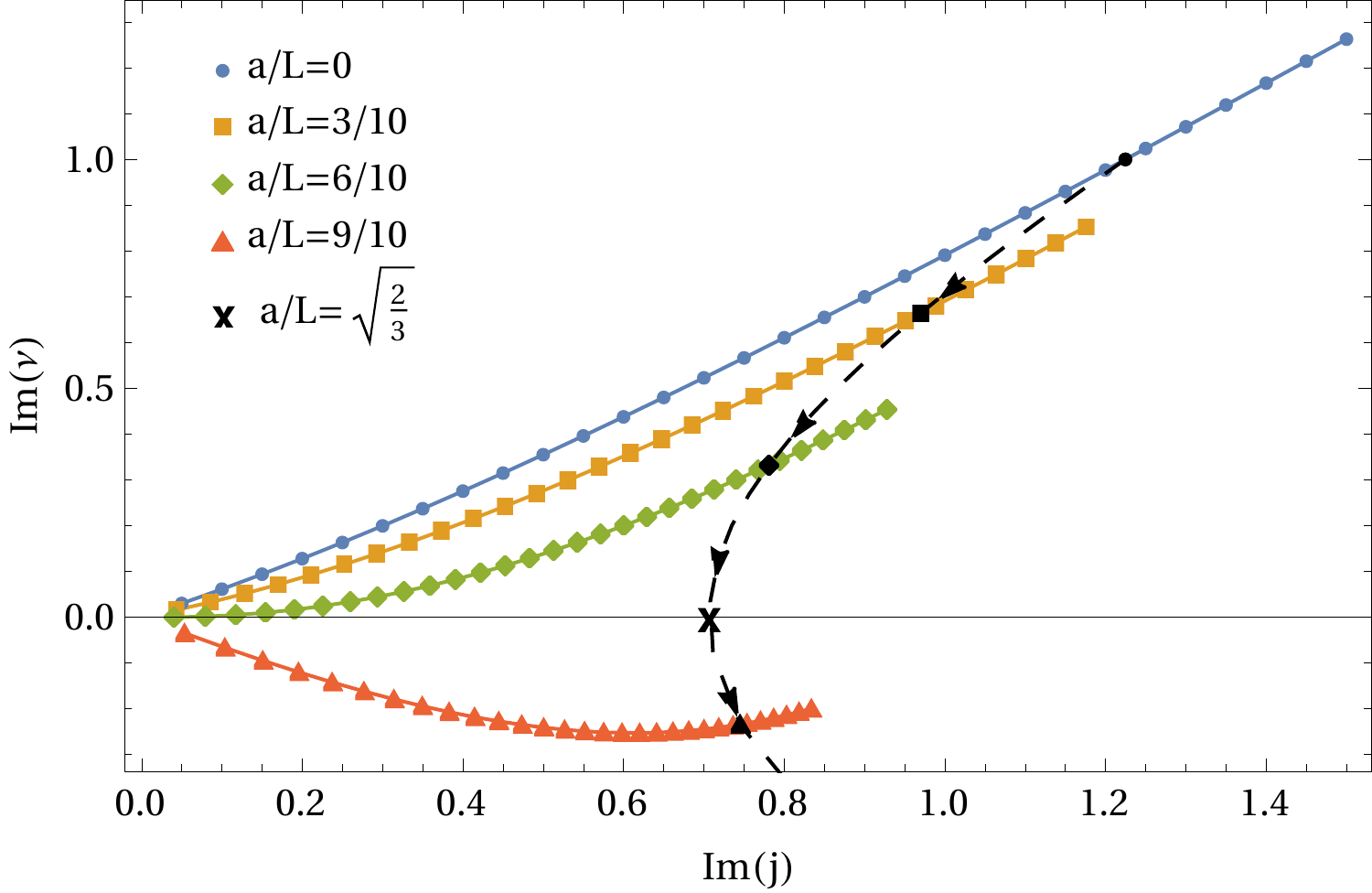}
    \caption{
        \textit{Chaos in the sound dispersion.}  
            The complex-valued dispersion relation $\nu(j)$ for the rotating black hole in the large black hole limit (as given by equation~\eqref{eq:largeBHLimit} with the top signs). Displayed is the imaginary part of the frequency parameter $\text{Im} \nu$ of the perturbation as a function of the imaginary part of its angular momentum $\text{Im}j$. 
            Each color (and different symbol) represents a different value for the angular momentum per mass $a/L$ in incremental steps of $3/10$. 
            The black line indicates the trajectory of the pole-skipping location as the angular momentum per mass, $a/L$,  varies.
            The cross, $\mathbf x$, highlights the point where one of the pole-skipping points associated to this dispersion relation ceases to be located in the upper half plane, corresponding to $a/L = \sqrt{\frac{2}{3}}$.
    \label{fig:dispersion} 
    }
\end{figure}

\paragraph{} These numerical results are illustrated in Figure~\ref{fig:dispersion}. The pole-skipping locations \eqref{eq:Scalar_Pole_Skipping} obtained from a boost are indicated with black symbols in Figure~\ref{fig:dispersion} for increasing values of rotation $a/L$. As we previously noted at the special value $a/L=\sqrt{2/3}$ at which the boost speed is equivalent to the conformal butterfly velocity, the pole-skipping point associated with $(\omega, {\cal K}) = (\omega_+, i  k_+)$ crosses over into the lower half plane of the complex frequency plane (which has been highlighted in the figure with a cross).\footnote{It would be interesting to study how rotation affects the relation between critical points and pole-skipping points for gapped modes found in~\cite{Abbasi:2020xli}. There, it was discovered that critical points computed from massive scalar, vector and tensor probe fields are bounded from above by the set of all pole-skipping points when the dimension of the dual operator (mass of the perturbed field on the gravity side) is treated as a parameter, see figure 8 in~\cite{Abbasi:2020xli}. Note, that massive (scalar, tensor, and vector) probe fields are considered in~\cite{Abbasi:2020xli}, whereas in the present work we consider metric perturbations; the equation of motion for the spin-2 (tensor) metric perturbation agrees with the massless tensor probe field of~\cite{Abbasi:2020xli}. 
}  

\subsubsection*{Lyapunov exponent and butterfly velocities}

\paragraph{} We have now explicitly confirmed the existence of pole-skipping in the energy density Green's functions at $(\omega, {\cal K})=(\omega_\pm, \pm i k_\pm)$, which precisely matches the functional form of the OTOC on Hopf circles in the large black hole limit we presented in Section~\ref{Sect_OTOC_Hopf}. It is now interesting to highlight several surprising physical implications of these results. 

\begin{figure}[htbp]
\begin{center}
    \includegraphics[width=12cm]{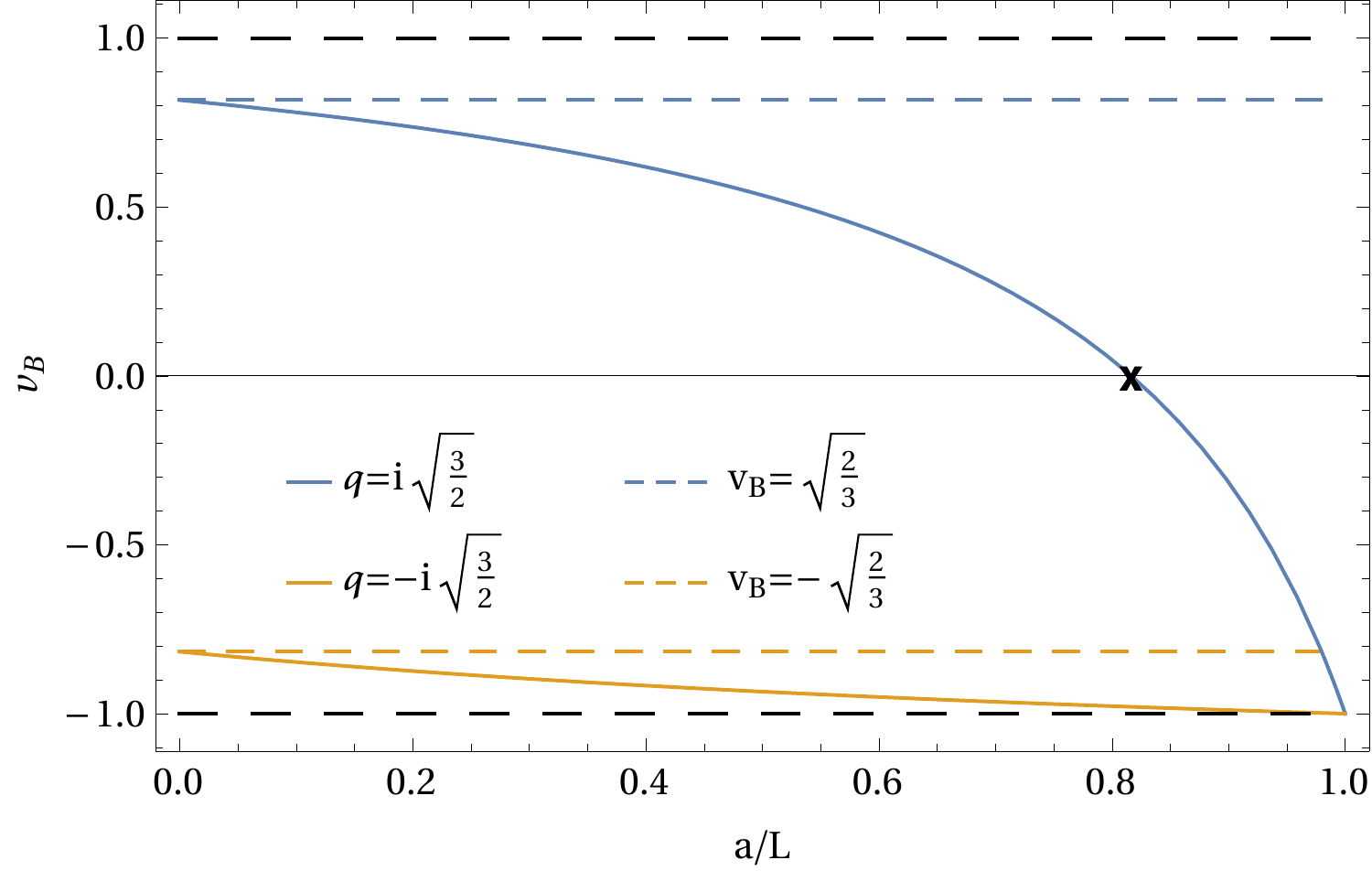}
\end{center}
    \caption{
    {\it Butterfly velocity in a rotating plasma.} The butterfly velocities $v_B^{\pm}$ given in \eqref{eq:vtildeB} are displayed as a function of the rotation parameter $a/L$.
    The colored dashed lines indicate the conformal value $v_B^{(0)}=\pm \sqrt{2/3}$ whilst the black dashed lines indicate the speed of light. 
    Notice that the ``upstream'' butterfly velocity $v_B^+$ crosses zero at exactly the same value of rotation $a/L = \sqrt{\frac{2}{3}}$ for which the pole-skipping point plotted in Figure~\ref{fig:dispersion} ceases to be located in the upper half complex frequency plane.
    \label{fig:butterfly}}
\end{figure}

\paragraph{} In particular, the fact that the pole-skipping location $(\omega, {\cal K}) = (\omega_+,i k_+)$ enters the lower half complex frequency plane when the boost associated to rotation exceeds $v_B^{(0)}$ has interesting implications for the associated butterfly velocity. In particular extracting a butterfly velocity from the pole-skipping locations (or equivalently the OTOC profile as we did in Section~\ref{Sect_OTOC_Hopf}) by $\omega_{\pm}/k_{\pm} = v_B^{\pm}$ we obtain associated butterfly velocities
\begin{equation}\label{eq:vtildeB}
    v_B^{\pm}=\frac{\sqrt{\frac{2}{3}}\mp \frac a L }{1\mp\sqrt{\frac{2}{3}} \frac a L} \, ,
\end{equation}
which is nothing but relativistic addition of velocities obtained by applying a boost with speed $a/L$ parallel to $\pm v_B^{(0)}$ \footnote{This has been independently verified in private communication between the authors, Casey Cartwright and Matthias Kaminski, with Navid Abbasi.}. The fact that this is an addition of velocities stems from the transformation given in equation (\ref{eq:criticalPointTrajectory}) being a Lorentz boost, as discussed in~\cite{Cartwright:2021qpp}. The butterfly velocity $v_B^+$ is plotted in Figure~\ref{fig:butterfly} as a function of the rotation parameter $a/L$. Note that the butterfly velocity $v_B^+$ vanishes at $a/L=\sqrt{2/3}$ (the magnitude of the conformal butterfly velocity) precisely at same time that the pole skipping location $(\omega, {\cal K}) = (\omega_+, i k_+)$ passes into the lower half complex frequency plane. 

\paragraph{}The vanishing of the butterfly velocity $v_B^+$ at $a/L=\sqrt{2/3}$ can be naturally interpreted with the help of the large black hole limit. In this limit, perturbations of the rotating black hole experience a fluid boosted with a velocity $a/L$. A perturbation traveling upstream against this fluid flow in a conformal theory will be experienced by an observer at rest as sitting still exactly when the perturbation has the speed $|a/L|$. Thus, a perturbation moving at the conformal butterfly speed $\sqrt{2/3}$ appears to sit still when the fluid streams with velocity $|a/L|=\sqrt{2/3}$. When the fluid flows faster, $|a/L|>\sqrt{2/3}$, this  perturbation is dragged downstream. This phenomenon could not be observed in previous studies of OTOCs and pole-skipping in rotating black holes. In particular, in the BTZ black hole case~\cite{Liu_2020,2019Jahnke} the butterfly velocity is equal to the speed of light, and likewise in studies of pole-skipping and OTOCs in Kerr-AdS \cite{blake2021chaos} were restricted to the slowly rotating limit in the which the rotation speed was parametrically smaller than the corresponding conformal butterfly velocity.  

\paragraph{} Finally, we note that the Lyapunov exponent $\lambda_L$ associated to the OTOC on Hopf circles in the large black hole limit can be extracted from $H(t, 0)$ in \eqref{hopfotoc} (see \cite{2020Mezei}). This gives $\lambda_L = 2 \pi T_+ = \mathrm{Im}(\omega_+)$ (for $a>0$) and $\lambda_L = 2 \pi T_- = \mathrm{Im}(\omega_-)$ (for $a<0$). We thus have 
\begin{equation}
 \lambda_L = 2\pi T \left(1-\sqrt{\frac{3}{2}}\frac{|a|}{L}\right) = 2 \pi T \left(1-|v|/v_B^{(0)} \right)
\end{equation}
which saturates a velocity-dependent generalization of the Maldacena/Shenker/Stanford chaos bound \cite{Maldacena:2015waa} that was proposed in~\cite{2020Mezei}. Furthermore, this saturates the generalized bound for the positive operators $\theta_a Q_a$  proposed in~\cite{2020Mezei}, given by
\begin{equation}\label{eq:gen_bound}
    \frac{\left| \theta_a[Q_a]^\mu \partial_\mu H \right|}{1-H}\leq 2\pi  \, .
\end{equation}
where $H$ is the OTOC, $\theta_a$ are chemical potentials and $Q_a$ are the corresponding generators. In our case the operators are given by $\theta_a Q_a=\beta(\tilde{H}+\Omega J)$ (with $\beta$ the inverse temperature, $\tilde{H}$ the Hamiltonian, $J$ the angular momentum and $\Omega$ the angular velocity) and the generalized bound given above in equation (\ref{eq:gen_bound}) reduces to
\begin{equation}
    \frac{\left| \partial_t H(t,\psi)+\Omega \partial_\psi H(t,\psi)\right|}{1-H(t,\psi)}\leq 2\pi T \, .
\end{equation}
Inserting the definition of the OTOC given in equation (\ref{otocsimple}) one finds the bound is saturated. 
It was pointed out already in~\cite{blake2021chaos}, that rotating systems in 1+1 d (e.g.~\cite{2019Jahnke}) have been noticed to satisfy the generalized MSS bound of~\cite{2020Mezei}.

%
\section{Discussion}
\label{sec:conclusion}

\paragraph{} We have studied the relationship between many-body quantum chaos and energy dynamics for the theory dual to a five-dimensional Myers-Perry-AdS black hole with equal angular momenta. In the context of many-body chaos we obtain a closed form expression~\eqref{exactshock} for the gravitational shock wave at the horizon that governs the functional form of the OTOC. Furthermore in Section~\ref{NH} we demonstrated using a near-horizon expansion that the retarded energy density response of the dual boundary theory exhibits pole-skipping at a frequency $\omega = i 2 \pi T$ (in co-rotating coordinates) whenever the angular profile of bulk fluctuations satisfies the (sourceless) shockwave equation at the horizon. Concretely, this condition corresponds to the constraint~\eqref{poleskippingW} on the quantum numbers ${\cal J}$, ${\cal K}$, ${\cal M}$ of Wigner D-functions that are typically used to diagonalise the angular response of the boundary theory on the 3-sphere. 

\paragraph{} Furthermore, we demonstrated that in the large black hole limit this leads to a precise connection between the functional form of the OTOC for operators lying on Hopf circles of the boundary $S^3$, and a family of pole-skipping points associated to perturbations whose angular momentum was aligned with the rotation of the background geometry (i.e. ${\cal K} = {\cal J}$). In particular, we found that there was pole-skipping in the dual rest frame at complex values of frequency and momenta that precisely match those extracted from the exponential profile of the OTOC on Hopf circles (see~\eqref{hopfotoc} and \eqref{eq:poleSkippingPointsRestFrame}), in an entirely analogous manner to the usual statements of pole-skipping in static planar black holes.

\paragraph{}It is worth emphasising that in contrast to previous studies of the Kerr-AdS black hole~\cite{blake2021chaos}, these results are valid for any value of the rotation parameter $a/L$. Furthermore, we were also able to explicitly confirm numerically that the dispersion relations of quasi-normal modes in the sound channel of scalar perturbations pass through the pole-skipping locations as expected. Intriguingly, we found that both the form of the OTOC on Hopf circles and the associated pole-skipping points with ${\cal K} = {\cal J}$ were, in the large black hole limit, equivalent to those of a boosted black brane. 

\paragraph{}Despite their simple form, these results for large black holes had several interesting implications for the physics of the dual theory. Firstly, as in previous examples of rotating black holes, whilst the pole-skipping points associated to chaos we examined always take place at a frequency $\omega = i 2 \pi T$ in co-rotating angular coordinates, the pole-skipping frequencies are shifted in the rest frame of the boundary theory. Interestingly, we found that for large black holes with $\cal K = \cal J$ one of these pole-skipping points passes into the lower half plane when the velocity $v = a/L$ associated to rotation exceeds the value of the butterfly velocity in the conformal non-rotating theory. Furthermore, at the same time, the butterfly velocity associated to ``upstream'' perturbations becomes negative. Whilst these features can be understood from the equivalence to a boosted black brane, this phenomenon could not be seen in previous results in rotating black holes in BTZ~\cite{2020Mezei} or Kerr-AdS black holes~\cite{blake2021chaos}. Furthermore, we found that the Lyapunov exponent associated to chaos saturated a generalised MSS bound for rotating ensembles proposed in \cite{2020Mezei}. It would be interesting to understand in the future how general these phenomena are for other examples of rotating black holes. 

\paragraph{} Finally, we emphasise that whilst the detailed analysis of Section~\ref{Num} was limited to perturbations of large black holes with ${\cal K} = {\cal J}$, our near horizon analysis predicts pole-skipping for any value of $r_0/L$ whenever the quantum numbers ${\cal J}, {\cal K}, {\cal M}$ characterising the angular dependence of quasi-normal modes satisfy the constraint~\eqref{poleskippingW}.  In principle, there are two types of corrections to the results of Section~\ref{Num}, both those associated to $L/r_0$-corrections and those from relaxing the condition $\mathcal{K}=\mathcal{J}$. The physical significance of $L/r_0$-corrections is that in the strict large black hole limit the perturbation equations for $\mathcal{K} = \mathcal{J}$ are identical to perturbation equations of a boosted black brane; and the boost is a symmetry transformation. When taking into account $L/r_0$-corrections, the perturbation equations for the rotating black hole are no longer related to a black hole at rest by any symmetry transformation we are aware of. From example computations of the $L/r_0$-corrections~\cite{Cartwright:2021qpp}, we expect effects on the pole-skipping locations to be smooth and still negligible at $r_0/L\approx \mathcal{O}(10)$. Thus, we expect our results to be a good approximation to the full rotating black hole result if the black hole is sufficiently large $r_0/L> 10$. More difficult and possibly more drastic could be the effect of allowing perturbations to rotate transverse to the direction of rotation of the fluid by choosing $\mathcal{K}\neq \mathcal{J}$, upon which the bulk metric perturbations do not (naively) decouple into distinct scalar, tensor and vector sectors. However, it would be interesting to test explicitly the existence of pole-skipping at the locations at ~\eqref{poleskippingW} with this condition relaxed. In principle, this should still be tractable due to the fact that the bulk perturbations can be decoupled using the Wigner D-functions into a series of coupled ODEs for the radial profiles $\delta h_{a b}(r)$ of metric perturbations, and we leave this as an interesting question for future work.

\acknowledgments
We are grateful to Richard Davison and Diandian Wang for very helpful discussions. This work
was supported, in part, by the U.S.~Department of Energy grant DE-SC-0012447. This work is supported, in part, by the Netherlands Organisation for Scientific Research (NWO) under the VICI grant VI.C.202.104. AT acknowledges support from UK EPSRC (EP/SO23607/1).

\appendix
\addcontentsline{toc}{section}{Appendices}
\renewcommand{\thesection}{\Alph{section}}

\section{Analytic solution to shock-wave equation}
\label{appendix:Solve_shock}

Here include the details of the computation of the analytic expression for the shock-wave profile for large black holes when $\theta = \theta' = 0$. Recall, the exact solution to \eqref{normalisedshock} for the shockwave profile $f(\tilde{\psi},  \phi, \theta, \theta')$ and with a normalised delta function source at $(\tilde{\psi}^{\prime}=0, \phi^{\prime}=0, \theta^{\prime})$ is given by \eqref{exactshock}, which for clarity we reproduce below 
\begin{equation}
f(\tilde{\psi}, \phi, \theta, \theta^{\prime}) =  \sum_{{\cal J}=0, 1/2, 1/, \dots}^{\infty} \sum_{{\cal K}=-{\cal J}}^{{\cal J}}  \sum_{{\cal M}=-{\cal J}}^{{\cal J}}     \frac{2{\cal J}+1}{16 \pi^2}  \frac{d^{{\cal J}}_{{\cal K} {\cal M}}(\theta^{\prime})  d^{{\cal J}}_{{\cal K} {\cal M}}(\theta)}{{\cal J}({\cal J}+1)- \lambda_1 - i \lambda_2 {\cal K} + \lambda_3 {\cal K}^2}  e^{i {\cal K} \tilde{\psi} + i {\cal M} \phi},  
\label{exactshockapp}
\end{equation}%
with $\lambda_1,  \lambda_2, \lambda_3$ the constants given in~\eqref{lambdas}. As discussed in the main text, the OTOC on Hopf circles in the large black hole limit can be extracted from the shockwave profile with $\theta = \theta' = 0$. In this case, it is possible to perform the infinite sum in \eqref{exactshockapp} analytically and obtain a simple expression for the OTOC. 
\paragraph{} In particular, using that $d^{\cal J}_{{\cal K} {\cal M}}(0) = \delta_{{\cal K} {\cal M}}$ (see e.g. \cite{Khersonskii:1988krb})
we have that \eqref{exactshockapp} reduces to  %
\begin{equation}\label{eq:reduced_ShockWave}
f(\alpha) =  \sum_{{\cal J}=0,1/2,1,\dots}^{\infty} \sum_{{\cal K}=-{\cal J}}^{{\cal J}}   \frac{2 {\cal J}+1}{16 \pi^2}  \frac{e^{i {\cal K} \alpha}}{{\cal J}({\cal J}+1)- \lambda_1 - i  \lambda_2 {\cal K} + \lambda_3 {\cal K}^2}, 
\end{equation}
where $\alpha = \tilde{\psi} +\phi$. The large black hole limit corresponds to taking ${\cal K} \sim {\cal J} \sim r_0/L \gg 1$ with ${\cal K} \alpha$ fixed. In this limit the summations in \eqref{eq:reduced_ShockWave} can be replaced by integrals to obtain
\begin{equation}
f(\alpha) =  \int_{0}^{\infty} d {\cal J} \int_{-{\cal J}}^{{\cal J}} d {\cal K}   \frac{{\cal J}}{4 \pi^2}  \frac{e^{i {\cal K} \alpha}}{{\cal J}^2- \lambda_1 - i  \lambda_2 {\cal K} + \lambda_3 {\cal K}^2} .\nonumber
\end{equation}
We now make the substitution ${\cal K} = {\cal J} \cos t$ where $t \in [0, \pi]$. The integral becomes
\begin{equation}
f(\alpha) = {\int_0}^{\infty} d {\cal J} \int_{0}^{\pi} dt  \frac{{\cal J}^2 \sin t}{4 \pi^2}  \frac{e^{i \alpha {\cal J} \cos t}}{{\cal J}^2 - \lambda_1 - i  \lambda_2 {\cal J} \cos t + \lambda_3 {\cal J}^2 \cos^2 t}, \nonumber
\end{equation}
which can equivalently be written as  
\begin{equation}
f(\alpha) =  \int_{0}^{\pi/2} dt\int_{-\infty}^{\infty} d {\cal J}   \frac{{\cal J}^2 \sin t}{4 \pi^2}  \frac{e^{i \alpha {\cal J} \cos t}}{{\cal J}^2 - \lambda_1 - i  \lambda_2 {\cal J} \cos t + \lambda_3 {\cal J}^2 \cos^2 t},
\label{integral3}
\end{equation}
where we have first adjusted the limits on the integrals and then swapped the order of integration.  The above expression is convenient because the $\cos t$ factor in the exponent is now manifestly positive for $t \in (0, \pi/ 2)$. The above integral can then be computed by using contour integration to first perform the ${\cal J}$ integral, and then finally performing the remaining integration over $t$. 
\paragraph{} In particular, we note that the integrand in \eqref{integral3} has simple poles at ${\cal J}_{\pm}(t)$ where
\begin{equation}
{\cal J}_{\pm}(t) = \frac{-2 i a r_0 \sqrt{1 - a^2/L^2} \cos t \pm i r_0 \sqrt{6 L^2 - 5 a^2 - a^2 \cos 2 t}}{2 L^2 - 2 a^2 \cos^2 t}
\label{poles}
\end{equation}
To evaluate the integral over ${\cal J}$ we then consider an integration contour consisting of the integral along the real ${\cal J}$ axis interval $[-R, R]$ and then along a semicircle $\Gamma_R$ in the UHP complex ${\cal J}$ plane where we set ${\cal J} = R e^{i \chi}$ for $0 \leq \chi \leq \pi$. Denoting the integral along the interval $[-R,R]$ as $I_R$ and the integral long the semicircle as $I_{\Gamma_R}$ we have from the residue theorem that $I_R + I_{\Gamma_R}$ is $2 \pi i$ times the residue at ${\cal J}_+(t)$ (which is the only pole in the upper half plane for $t \in (0, \pi/2)$). This gives 
\begin{equation}
I_R + I_{\Gamma_R} =  -\int_{0}^{\pi/2} dt \frac{i}{2 \pi} \frac{{\cal J}_+^2(t) \sin t}{(1 - a^2/L^2 \cos^2 t)({\cal J}_+(t)- {\cal J}_-(t))} e^{i \alpha {\cal J}_+(t) \cos t}.
\label{tintegral}
\end{equation}
In order to perform the $t$ integral we now note that using the explicit expressions for ${\cal J}_{\pm}(t)$ in \eqref{poles} we find the identity, 
\begin{equation}
\frac{{\cal J}_+^2(t) \sin t}{(1 - a^2/L^2 \cos^2 t)({\cal J}_+(t)- {\cal J}_-(t))} = -\frac{1}{2} \frac{d({\cal J}_+(t) \cos t)}{dt},   \nonumber 
\end{equation}
from which we deduce that the integrand in \eqref{tintegral} is a total derivative. Performing the $t$ integral we therefore obtain 
\begin{equation}
I_R + I_{\Gamma_R} =  \frac{1}{4 \pi \alpha} \bigg[ 1 - e^{- {\cal K}_+ \alpha} \bigg], \nonumber
\end{equation}
where ${\cal J}_+(0) = i {\cal K}_+$ with
\begin{equation}
{\cal K}_+ = \frac{r_0}{L} \frac{1}{\sqrt{1 - a^2/L^2}}\bigg(\sqrt{\frac{3}{2}} - \frac{a}{L}\bigg) = k_+.
\end{equation}
The shockwave profile for $\alpha > 0$ is then given by $f(\alpha) = \lim_{R \to \infty} I_{R}$. To obtain this, we note that using the standard parameterisation ${\cal J} = R e^{i \chi}$ on $\Gamma_R$ it is straightforward to show that $\lim_{R \to \infty} I_{\Gamma_{R}}= 1/(4 \pi  \alpha)$. We thus obtain the shockwave profile 
\begin{equation}
f(\alpha) = -\frac{e^{- {\cal K}_+ \alpha}}{4 \pi \alpha}, \;\;\;\;\;\;\;\;\;\ \alpha > 0. 
\end{equation}
An entirely analogous computation can be performed for the case $\alpha < 0$ by considering a semi-circular contour in the lower half ${\cal J}$ plane. Following the same steps outlined above we find 
\begin{equation}
f(\alpha) = \frac{e^{{\cal K}_- \alpha}}{4 \pi \alpha}, \;\;\;\;\;\;\;\;\;\ \alpha < 0,
\end{equation}
where 
\begin{equation}
{\cal K}_-  = \frac{r_0}{L} \frac{1}{\sqrt{1 - a^2/L^2}}\bigg(\sqrt{\frac{3}{2}} + \frac{a}{L}\bigg) = k_{-}. 
\end{equation}
%


\section{Pole-skipping in other channels}
\label{appendix:otherchannels}

\paragraph{} As discussed in Section~\ref{Num}, the gravitational perturbation equations of the AdS Myers-Perry black hole with quantum numbers satisfying ${\cal K} = {\cal J}$ dramatically simplify in the large black hole limit, and are equivalent to those of a boosted black brane. One can thus obtain pole-skipping locations by applying a boost with velocity $v = a/L$ to the pole-skipping locations of the static Schwarzschild AdS$_5$ black brane. In the main text we discussed this in the context of the pole skipping locations associated to many-body quantum chaos, i.e those that take place in the scalar channel at a frequency $\omega = i 2 \pi T$ in co-rotating coordinates. However it is well known that there exist other pole skipping locations, in the lower half complex frequency plane \cite{Grozdanov_2019, 2020}, both in the scalar sector and in other channels. Whilst not related directly to quantum many-body chaos, these other pole-skipping locations provide important constraints on the spectrum on QNMs of a theory.

\paragraph{} As an illustration, we now utilize \eqref{eq:physicalBoost} further to identify examples of pole-skipping locations in the tensor and vector sectors of the AdS Myers-Perry black hole. Explicit forms of the tensor and vector fluctuation equations can be found in~\cite{Garbiso:2020puw} and~\cite{Cartwright:2021qpp}, respectively. 
For a Schwarzschild-AdS$_5$ black brane at rest in the spin 1 and spin 2 sector the pole-skipping points are given by, in the notation of Section~\ref{Num}, as \cite{Grozdanov_2019}
\begin{subequations}
    \begin{align}\label{eq:vectorTensorPoleSkipping}
          \text{Spin 1 - Vector:} \hspace{1cm}   \mathfrak{w} &= -i,\hspace{1cm} \mathfrak{q} = \pm  \sqrt{\frac{3}{2}}   \, , \nonumber \\
        \text{Spin 2 - Tensor:}  \hspace{1cm}     \mathfrak{w} &= -i, \hspace{1cm} \mathfrak{q}= \pm  i \sqrt{\frac{3}{2}}  \, . \nonumber
    \end{align}
\end{subequations}
Using the transformation in \eqref{eq:criticalPointTrajectory} one immediately obtains
\begin{subequations}
    \begin{align}
              \text{Spin 1 - Vector:} \hspace{1cm}    \nu_{vector}&= -\frac{\pm\sqrt{3} a+ i\sqrt{2}}{\sqrt{2}\sqrt{1-a^2}}\,,\hspace{1cm} j_{vector}= \frac{\pm\sqrt{3}+ i a\sqrt{2}}{\sqrt{2}\sqrt{1-a^2}} \, , \nonumber \\
        \text{Spin 2 - Tensor:}  \hspace{1cm}    \nu_{tensor} &=-\frac{\pm\sqrt{3} a + \sqrt{2}}{\sqrt{2}\sqrt{1-a^2}}i\,, \hspace{1cm}  j_{tensor}=\frac{\pm\sqrt{3}+  a\sqrt{2}}{\sqrt{2}\sqrt{1-a^2}} i  \, . \nonumber
    \end{align}
\end{subequations}
where these results are valid at leading order in the large black hole limit for perturbations with ${\cal K} = {\cal J}$.

\section{Details of the numerical method}
\label{appendix:numericalConsistency}
\paragraph{} The linearized Einstein equations for gravitational perturbations take the form~\cite{Wald:1984rg}
\begin{equation}\label{eq:pertgenericeom}
    \dot{R}_{\mu\nu} = \frac{2\Lambda}{D-2}\delta g_{\mu\nu}\,\nonumber
\end{equation}
where
\begin{equation}
    \dot{R}_{\mu\nu} = -\frac{1}{2}\nabla_\mu \nabla_\nu \delta g-\frac{1}{2}\nabla^\lambda \nabla_\lambda \delta g_{\mu\nu}+\nabla^\lambda \nabla_{(\mu}\delta g_{\nu)\lambda}\,,\nonumber
\end{equation}
and $\Lambda = -6/L^2$, $D=5$ and $\delta g=\delta g^{~~\mu}_{\mu}=\delta g_{\nu \mu} g^{\mu \nu}$.  
The covariant derivatives are defined with respect to the background metric~\eqref{metric}. 
The rotating black hole metric~\eqref{metric} has a spatial $SU(2)\times U(1)$ symmetry along with a time translation symmetry which is reflected in a set of 5 Killing vectors. Construction of a mutually commuting set of observables leads to the decomposition of the metric in terms of Wigner D-functions, as discussed in Section~\ref{Num}. Using this decomposition in the linearized Einstein equation leads to a set of coupled ODEs for the radial profiles $h_{ab}(r)$ as discussed in Section~\ref{Num} and references therein. As discussed in Section~\ref{Num} upon setting setting ${\cal K} = {\cal J}$ the equations for $h_{ab}(r)$ decouple into distinct scalar, vector and tensor sectors. The relevant components of $h_{ab}(r)$ for the scalar sector are displayed in~\cite{Garbiso:2020puw}. Taking the large black hole limit in \eqref{eq:largeBHLimit} and \eqref{eq:largeBHLimitCoor} the fluctuation equations reduce to those of a Schwarzschild black brane uniformly boosted along one of the spatial directions~\cite{Garbiso:2020puw,Cartwright:2021qpp}. 

\paragraph{} The quasi-normal mode frequencies of the dual theory can by extracted from the system of coupled ODEs for $h_{ab}(r)$  for each sector in one of two ways. Firstly one can introduce a gauge invariant master field variable as discussed in Section~\ref{Num}, reducing the the set of coupled equations to a single ODE. Alternatively one can represent the coupled set of ODEs for the relevant modes $h_{ab}(r)$ as a generalized eigenvalue problem. In this case for performing our numerics of the QNM dispersion relations in the scalar sector we have chosen to obtain the QNM frequencies by solving the generalized eigenvalue problem~\cite{Jansen:2017oag}. Schematically, the linearized Einstein field equations takes the form, $A\cdot\phi=\nu B\cdot \phi$, where $\phi$ represents a set of coupled fields in the fluctuation equations and $A$ and $B$ are differential operators, which depend on the value of the momentum $j$ (but do not depend on the frequency $\nu$). Numerically, the operators are represented as discrete differential operators by invoking an $N$th order truncated Chebyshev representation of the fields. One then solves the resulting linear system for the quasi-normal mode frequency $\nu$ given a choice of momentum $j$. The discretization procedure introduces spurious modes, which can be filtered by comparing the frequencies obtained when using an $N$th order Chebyshev representation of the fields to that of an $M$th order Chebyshev representation (with $M>N$). 

\paragraph{} The data displayed in Figure~\ref{fig:dispersion} is the result of solving the generalized eigenvalue problem associated with scalar perturbations of the static Schwarzschild black brane. The obtained dispersion relation, as displayed in blue, is then boosted according to \eqref{eq:physicalBoost} to obtain the orange, green and red curves. The computed modes are represented as dots while the lines are simply interpolations of the data, added to the plot to guide the eye. 
\begin{figure}[t]
    \centering
    \includegraphics[width=12cm]{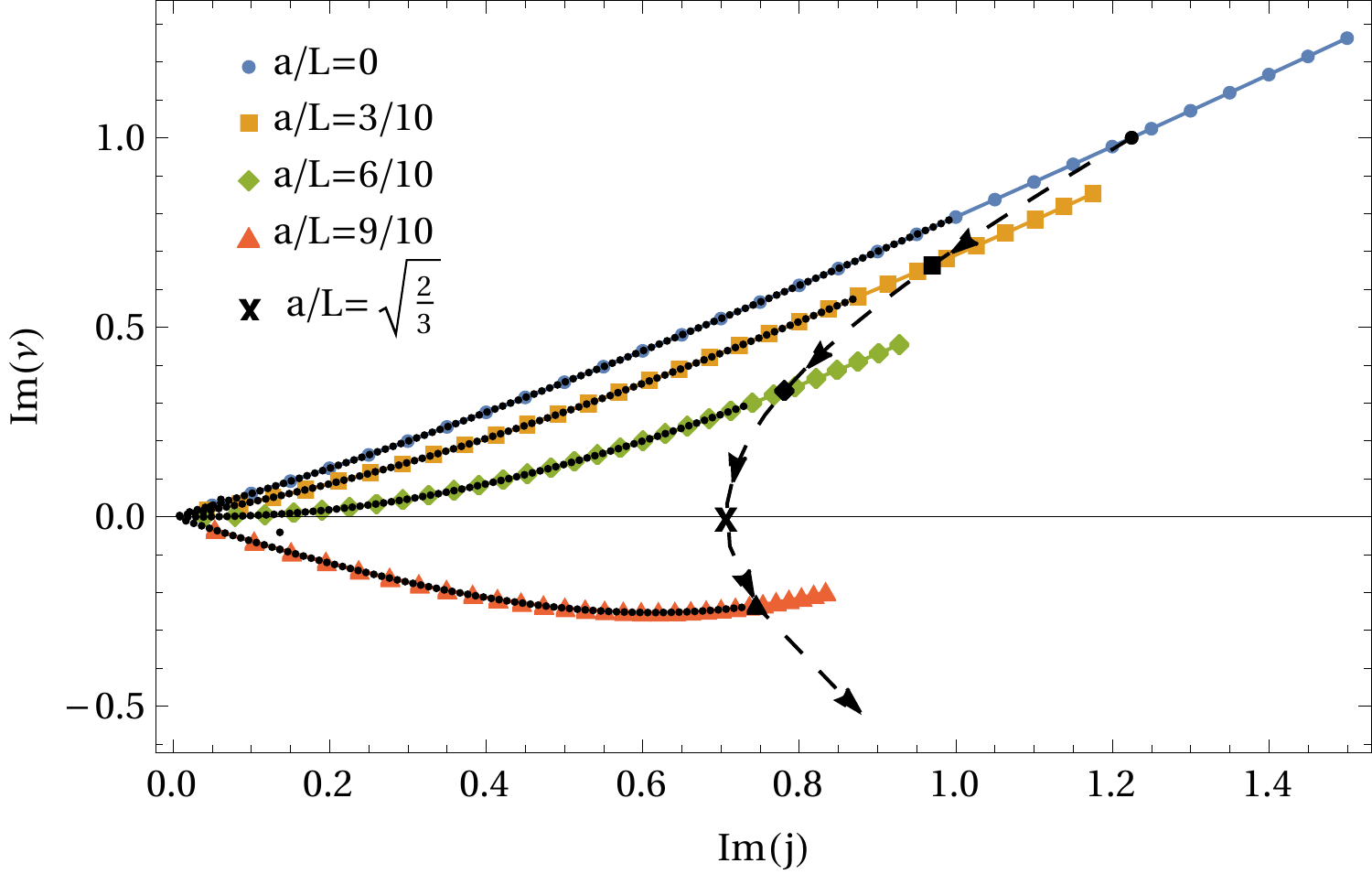}
    \caption{
        \textit{More chaos in the sound dispersion.} 
            The dispersion relation $\nu(j)$ for the rotating black hole in the large black hole limit~(\ref{eq:largeBHLimit}) as already displayed in Figure~\ref{fig:dispersion}. Here, 
the black dots were added as they were computed independent of the colored data points. The colored data points were calculated in a static Schwarzschild black brane at rest with subsequently applying the boost transformation~(\ref{eq:criticalPointTrajectory}). The black dots were calculated directly from the perturbation equations of the rotating black hole in the large black hole limit. These two methods are in excellent agreement with the difference between the data for each point within the parameter range displayed here being on the order of $10^{-6}$. 
    \label{fig:dispersionPlusSecondMethod} 
    }
\end{figure}

\paragraph{} As a check of both our numerics and our understanding of the interpretation of the large black hole limit, we have computed the dispersion relations displayed in Figure~\ref{fig:dispersion} by a second independent numerical method using an independently constructed routine. The second routine works directly with the equations describing the scalar sector of perturbations of the rotating black hole in the large black hole limit (i.e. no boost transformations). The results of the second approach are displayed as small black dots in Figure~\ref{fig:dispersionPlusSecondMethod} along with the modes, shown in color, as computed in the first approach. Interpolating the mode frequencies as a function of the momentum, we have compared the relative percent difference, $\mathfrak{r}\equiv 2(A-B)/(A+B)$ of the two independent codes and found $\mathfrak{r}\sim O(10^{-6})$ for every point displayed. The agreement of these two independent schemes provides a consistency check on the numerical results and gives an independent verification of the interpretation of the large black hole limit of the Myers-Perry-AdS black hole as a boosted Schwarzschild black brane.


\bibliographystyle{JHEP}
\bibliography{bib}

\providecommand{\href}[2]{#2}\begingroup\raggedright\begin{thebibliography}{10}

\bibitem{2018G}
S.~Grozdanov, K.~Schalm and V.~Scopelliti, \emph{Black hole scrambling from
  hydrodynamics},
  \href{https://doi.org/10.1103/physrevlett.120.231601}{\emph{Physical Review
  Letters} {\bfseries 120} (2018) }.

\bibitem{2018B}
M.~Blake, H.~Lee and H.~Liu, \emph{A quantum hydrodynamical description for
  scrambling and many-body chaos},
  \href{https://doi.org/10.1007/jhep10(2018)127}{\emph{Journal of High Energy
  Physics} {\bfseries 2018} (2018) }.

\bibitem{2018}
M.~Blake, R.~A. Davison, S.~Grozdanov and H.~Liu, \emph{Many-body chaos and
  energy dynamics in holography},
  \href{https://doi.org/10.1007/jhep10(2018)035}{\emph{Journal of High Energy
  Physics} {\bfseries 2018} (2018) }.

\bibitem{2020}
M.~Blake, R.~A. Davison and D.~Vegh, \emph{Horizon constraints on holographic
  green’s functions},
  \href{https://doi.org/10.1007/jhep01(2020)077}{\emph{Journal of High Energy
  Physics} {\bfseries 2020} (2020) }.

\bibitem{Grozdanov_2019}
S.~Grozdanov, P.~K. Kovtun, A.~O. Starinets and P.~Tadi\'c, \emph{The complex
  life of hydrodynamic modes},
  \href{https://doi.org/10.1007/JHEP11(2019)097}{\emph{JHEP} {\bfseries 11}
  (2019) 097} [\href{https://arxiv.org/abs/1904.12862}{{\ttfamily
  1904.12862}}].

\bibitem{Blake:2021wqj}
M.~Blake and H.~Liu, \emph{{On systems of maximal quantum chaos}},
  \href{https://doi.org/10.1007/JHEP05(2021)229}{\emph{JHEP} {\bfseries 05}
  (2021) 229} [\href{https://arxiv.org/abs/2102.11294}{{\ttfamily
  2102.11294}}].

\bibitem{Grozdanov2_2019}
S.~Grozdanov, \emph{On the connection between hydrodynamics and quantum chaos
  in holographic theories with stringy corrections},
  \href{https://doi.org/10.1007/JHEP01(2019)048}{\emph{JHEP} {\bfseries 01}
  (2019) 048} [\href{https://arxiv.org/abs/1811.09641}{{\ttfamily
  1811.09641}}].

\bibitem{Natsuume_2020}
M.~Natsuume and T.~Okamura, \emph{Holographic chaos, pole-skipping, and
  regularity}, \href{https://doi.org/10.1093/ptep/ptz155}{\emph{PTEP}
  {\bfseries 2020} (2020) 013B07}
  [\href{https://arxiv.org/abs/1905.12014}{{\ttfamily 1905.12014}}].

\bibitem{Wang:2022mcq}
D.~Wang and Z.-Y. Wang, \emph{{Pole-skipping in holographic theories with
  bosonic fields}},  \href{https://arxiv.org/abs/2208.01047}{{\ttfamily
  2208.01047}}.

\bibitem{Liu_2020}
Y.~Liu and A.~Raju, \emph{Quantum chaos in topologically massive gravity},
  \href{https://doi.org/10.1007/JHEP12(2020)027}{\emph{JHEP} {\bfseries 12}
  (2020) 027} [\href{https://arxiv.org/abs/2005.08508}{{\ttfamily
  2005.08508}}].

\bibitem{2020Mezei}
M.~Mezei and G.~S\'arosi, \emph{{Chaos in the butterfly cone}},
  \href{https://doi.org/10.1007/JHEP01(2020)186}{\emph{JHEP} {\bfseries 01}
  (2020) 186} [\href{https://arxiv.org/abs/1908.03574}{{\ttfamily
  1908.03574}}].

\bibitem{2019Jahnke}
V.~Jahnke, K.-Y. Kim and J.~Yoon, \emph{On the chaos bound in rotating black
  holes}, \href{https://doi.org/10.1007/jhep05(2019)037}{\emph{Journal of High
  Energy Physics} {\bfseries 2019} (2019) }.

\bibitem{blake2021chaos}
M.~Blake and R.~A. Davison, \emph{Chaos and pole-skipping in rotating black
  holes},  \href{https://arxiv.org/abs/2111.11093}{{\ttfamily 2111.11093}}.

\bibitem{myers_perry_1986}
R.~C. Myers and M.~J. Perry, \emph{Black holes in higher dimensional
  space-times},
  \href{https://doi.org/10.1016/0003-4916(86)90014-x}{\emph{Annals of Physics}
  {\bfseries 171} (1986) 491}.

\bibitem{Malvimat:2022fhd}
V.~Malvimat and R.~R. Poojary, \emph{{Fast Scrambling of mutual information in
  Kerr-AdS$_{\textbf{5}}$}},
  \href{https://arxiv.org/abs/2210.02950}{{\ttfamily 2210.02950}}.

\bibitem{shenker2015stringy}
S.~H. Shenker and D.~Stanford, \emph{Stringy effects in scrambling},  2015.

\bibitem{Chen:2006xh}
W.~Chen, H.~Lu and C.~N. Pope, \emph{General kerr-nut-ads metrics in all
  dimensions}, \href{https://doi.org/10.1088/0264-9381/23/17/013}{\emph{Class.
  Quant. Grav.} {\bfseries 23} (2006) 5323}
  [\href{https://arxiv.org/abs/hep-th/0604125}{{\ttfamily hep-th/0604125}}].

\bibitem{Gibbons:2004js}
G.~W. Gibbons, H.~Lu, D.~N. Page and C.~N. Pope, \emph{Rotating black holes in
  higher dimensions with a cosmological constant},
  \href{https://doi.org/10.1103/PhysRevLett.93.171102}{\emph{Phys. Rev. Lett.}
  {\bfseries 93} (2004) 171102}
  [\href{https://arxiv.org/abs/hep-th/0409155}{{\ttfamily hep-th/0409155}}].

\bibitem{Murata:2008xr}
K.~Murata, \emph{{Instabilities of Kerr-AdS(5) x S**5 Spacetime}},
  \href{https://doi.org/10.1143/PTP.121.1099}{\emph{Prog. Theor. Phys.}
  {\bfseries 121} (2009) 1099}
  [\href{https://arxiv.org/abs/0812.0718}{{\ttfamily 0812.0718}}].

\bibitem{Hawking:1998kw}
S.~W. Hawking, C.~J. Hunter and M.~Taylor, \emph{{Rotation and the AdS / CFT
  correspondence}},
  \href{https://doi.org/10.1103/PhysRevD.59.064005}{\emph{Phys. Rev. D}
  {\bfseries 59} (1999) 064005}
  [\href{https://arxiv.org/abs/hep-th/9811056}{{\ttfamily hep-th/9811056}}].

\bibitem{Gibbons:2004ai}
G.~W. Gibbons, M.~J. Perry and C.~N. Pope, \emph{{The First law of
  thermodynamics for Kerr-anti-de Sitter black holes}},
  \href{https://doi.org/10.1088/0264-9381/22/9/002}{\emph{Class. Quant. Grav.}
  {\bfseries 22} (2005) 1503}
  [\href{https://arxiv.org/abs/hep-th/0408217}{{\ttfamily hep-th/0408217}}].

\bibitem{Khersonskii:1988krb}
V.~K. Khersonskii, A.~N. Moskalev and D.~A. Varshalovich, \emph{Quantum Theory
  Of Angular Momentum}. World Scientific Publishing Company, 1988,
  \href{https://doi.org/10.1142/0270}{10.1142/0270}.

\bibitem{Murata:2007gv}
K.~Murata and J.~Soda, \emph{A note on separability of field equations in
  myers-perry spacetimes},
  \href{https://doi.org/10.1088/0264-9381/25/3/035006}{\emph{Class. Quant.
  Grav.} {\bfseries 25} (2008) 035006}
  [\href{https://arxiv.org/abs/0710.0221}{{\ttfamily 0710.0221}}].

\bibitem{Murata:2008yx}
K.~Murata and J.~Soda, \emph{Stability of five-dimensional myers-perry black
  holes with equal angular momenta},
  \href{https://doi.org/10.1143/PTP.120.561}{\emph{Prog. Theor. Phys.}
  {\bfseries 120} (2008) 561}
  [\href{https://arxiv.org/abs/0803.1371}{{\ttfamily 0803.1371}}].

\bibitem{Garbiso:2020puw}
M.~Garbiso and M.~Kaminski, \emph{Hydrodynamics of simply spinning black holes
  \& hydrodynamics for spinning quantum fluids},
  \href{https://doi.org/10.1007/JHEP12(2020)112}{\emph{JHEP} {\bfseries 12}
  (2020) 112} [\href{https://arxiv.org/abs/2007.04345}{{\ttfamily
  2007.04345}}].

\bibitem{Cartwright:2021qpp}
C.~Cartwright, M.~G. Amano, M.~Kaminski, J.~Noronha and E.~Speranza,
  \emph{Convergence of hydrodynamics in rapidly spinning strongly coupled
  plasma},  \href{https://arxiv.org/abs/2112.10781}{{\ttfamily 2112.10781}}.

\bibitem{Kovtun:2005ev}
P.~K. Kovtun and A.~O. Starinets, \emph{Quasinormal modes and holography},
  \href{https://doi.org/10.1103/PhysRevD.72.086009}{\emph{Phys. Rev. D}
  {\bfseries 72} (2005) 086009}
  [\href{https://arxiv.org/abs/hep-th/0506184}{{\ttfamily hep-th/0506184}}].

\bibitem{Abbasi:2020xli}
N.~Abbasi and M.~Kaminski, \emph{{Constraints on quasinormal modes and bounds
  for critical points from pole-skipping}},
  \href{https://doi.org/10.1007/JHEP03(2021)265}{\emph{JHEP} {\bfseries 03}
  (2021) 265} [\href{https://arxiv.org/abs/2012.15820}{{\ttfamily
  2012.15820}}].

\bibitem{Maldacena:2015waa}
J.~Maldacena, S.~H. Shenker and D.~Stanford, \emph{A bound on chaos},
  \href{https://doi.org/10.1007/JHEP08(2016)106}{\emph{JHEP} {\bfseries 08}
  (2016) 106} [\href{https://arxiv.org/abs/1503.01409}{{\ttfamily
  1503.01409}}].

\bibitem{Wald:1984rg}
R.~M. Wald, \emph{General Relativity}. Chicago Univ. Pr., Chicago, USA, 1984,
  \href{https://doi.org/10.7208/chicago/9780226870373.001.0001}{10.7208/chicago/9780226870373.001.0001}.

\bibitem{Jansen:2017oag}
A.~Jansen, \emph{Overdamped modes in schwarzschild-de sitter and a mathematica
  package for the numerical computation of quasinormal modes},
  \href{https://doi.org/10.1140/epjp/i2017-11825-9}{\emph{Eur. Phys. J. Plus}
  {\bfseries 132} (2017) 546}
  [\href{https://arxiv.org/abs/1709.09178}{{\ttfamily 1709.09178}}].

\end{thebibliography}\endgroup

\end{document}